\documentclass[conference]{IEEEtran}
\IEEEoverridecommandlockouts
\usepackage{cite}
\usepackage{amsmath, amssymb, amsfonts}
\usepackage{algorithm}
\usepackage{algpseudocode}
\usepackage{graphicx}
\usepackage[font=small]{caption}
\usepackage{subcaption}
\usepackage{diagbox}
\usepackage{mathtools}
\usepackage{booktabs}
\usepackage{textcomp}
\usepackage{xcolor}
\usepackage{url}
\usepackage[colorlinks]{hyperref}

\def\BibTeX{{\rm B\kern-.05em{\sc i\kern-.025em b}\kern-.08em
    T\kern-.1667em\lower.7ex\hbox{E}\kern-.125emX}}

\DeclarePairedDelimiter\ceil{\lceil}{\rceil}
\DeclarePairedDelimiter\floor{\lfloor}{\rfloor}

\newcommand{\lineautorefname }{Line~}

\newcounter{DaveCommentCounter}
\newcounter{LimingCommentCounter}
\usepackage{censor} 
\StopCensoring

\begin{document}

\title{Uncovering the Metaverse within Everyday Environments: a Coarse-to-Fine Approach\IEEEauthorrefmark{3}
}

\author{\IEEEauthorblockN{Liming Xu\IEEEauthorrefmark{1}
}
\IEEEauthorblockA{\textit{Department of Engineering} \\
\textit{University of Cambridge}\\
Cambridge, United Kingdom \\
lx249@cam.ac.uk}\\
\IEEEauthorblockN{Andrew P. French}
\IEEEauthorblockA{\textit{School of Computer Science} \\
\textit{University of Nottingham}\\
Nottingham, United Kingdom \\
andrew.p.french@nottingham.ac.uk}\\
\and
\IEEEauthorblockN{Dave Towey\IEEEauthorrefmark{2}}
\IEEEauthorblockA{\textit{School of Computer Science} \\
\textit{University of Nottingham Ningbo China}\\
Ningbo, China \\
dave.towey@nottingham.edu.cn}\\
\IEEEauthorblockN{Steve Benford}
\IEEEauthorblockA{\textit{School of Computer Science} \\
\textit{University of Nottingham}\\
Nottingham, United Kingdom \\
steve.benford@nottingham.ac.uk}

\thanks{\IEEEauthorrefmark{1}This work was completed while the first author was a PhD student at University of Nottingham Ningbo China.}
\thanks{\IEEEauthorrefmark{2}Corresponding author.}
\thanks{\IEEEauthorrefmark{3}This paper has been accepted by The 48th IEEE International Conference on Computers, Software, and Applications (COMPSAC 2024) for publication.}
}

\maketitle

\begin{abstract}
The recent release of the Apple Vision Pro has reignited interest in the metaverse, showcasing the intensified efforts of technology giants in developing platforms and devices to facilitate its growth.
As the metaverse continues to proliferate, it is foreseeable that everyday environments will become increasingly saturated with its presence.
Consequently, uncovering links to these metaverse items will be a crucial first step to interacting with this new augmented world. 
In this paper, we address the problem of establishing connections with virtual worlds within everyday environments, especially those that are not readily discernible through direct visual inspection.
We introduce a vision-based approach leveraging Artcode visual markers to uncover hidden metaverse links embedded in our ambient surroundings.
This approach progressively localises the access points to the metaverse, transitioning from coarse to fine localisation, thus facilitating an exploratory interaction process. 
Detailed experiments are conducted to study the performance of the proposed approach, demonstrating its effectiveness in Artcode localisation and enabling new interaction opportunities.
\end{abstract}

\begin{IEEEkeywords} 
Metaverse, Interaction, Coarse-to-fine, Artcode, Localisation
\end{IEEEkeywords}

\section{Introduction}\label{sec:introduction}
People have been increasingly engaging in virtual activities (online meetings, augmented reality (AR) and virtual reality (VR) games, etc.) since the COVID-19 pandemic. 
This form of AR/VR enabled interactions may continue to grow \cite{xu2022connecting, wang2022survey}.
Moreover, the very recent release of the Apple Vision Pro \cite{apple2023vision} has reignited interest in the metaverse, showcasing the intensified efforts of technology giants in developing platforms and devices to facilitate metaverse interaction.

With the metaverse becoming ubiquitous, it is foreseeable that our everyday environments will be augmented by various forms of metaverse interactions.
As described by Xu et al. \cite{xu2017recognizing}, interaction occurs after recognising the presence of access points to the metaverse associated with everyday objects. 
While the interaction {\it affordances}, referring to the potential actions individuals perceive when interacting with objects in their environment, of noticeable (and consequently obtrusive) access points such as QR codes or text instructions are already established, their widespread integration into our everyday environments may detract from the environment's aesthetic appeal.
Therefore, in this paper, we investigate the problem of enabling interaction affordances through geometrically flexible, aesthetically pleasing, and unobtrusive access points.

Artcodes \cite{meese2013codes, benford2016accountable} provide a solution. 
These markers are both interpretable by humans and readable by machines, making them ideal candidates for creating such access points \cite{xu2022connecting}.
However, their interaction affordances are not inherently apparent due to their flexible forms and capability to blend in or ``camouflage'' as common visual patterns. 
This paper thus aims to bridge this gap by proposing a vision-based approach to enhance the visibility and intuitiveness of interaction affordances associated with Artcodes in everyday environments.

The proposed approach consists of two stages: coarse localisation, which generates a heatmap indicating the general presence of Artcodes, and fine localisation, which precisely identifies their exact locations. 
Indeed, a map showing the presence of access points to virtual worlds (the metaverse) would be the first step to interacting with these virtual worlds \cite{xu2022connecting}. 
People would then decide to approach these entrances and ``enter'' into the corresponding metaverse through further operations, such as scanning, if interested. 
This coarse-to-fine approach may thus enable an exploratory interaction process, wherein people progressively discover the metaverse in the physical world, thereby creating a rich interaction experience and opening up new design opportunities, such as serendipity \cite{andre2009discovery}.

While Xu et al. \cite{xu2022connecting} do not provide technical solutions for enabling such interaction, this work presents a concrete approach to achieving this goal.
It contributes the connection layer of the classic three-layer metaverse conceptual model \cite{benford2021metaverse, radoff2021metaverse}, with the other two layers being the physical world and the virtual world. 
In particular, it develops connections to the metaverse for scenarios that require aesthetic awareness, such as galleries; and in public environments, such as corridors and hallways where surface patterns are common decorations.

The main contributions of this paper are twofold:
\begin{itemize}
    \item We propose a coarse-to-fine vision-based approach that could be used for uncovering metaverse connections in our everyday environment. 
    \item We create a dataset and conduct detailed experiments to evaluate the proposed approach, illustrating its effectiveness to enable exploratory interaction. 
\end{itemize}

The rest of this paper is structured as follows. 
\autoref{sec:background} briefly introduces the Artcode approach and reviews the related work on visual markers and the metaverse. 
\autoref{sec:approach} presents the proposed approach.
\autoref{sec:experiments} details experimental studies on the proposed approach. 
\autoref{sec:discussion} examines the experimenttal results and discuss the limitations of this work. 
Finally, \autoref{sec:conclusion} concludes this paper and describes future work.

\section{Background}
\label{sec:background}
This section describes the Artcode approach, a topological visual-marker-based AR technique for augmenting everyday objects. 
This section also discusses related work on visual markers and connections to the metaverse.

\begin{figure}[t]
    \centering
    \begin{subfigure}[t]{0.09\textwidth}
        \includegraphics[width=\textwidth]{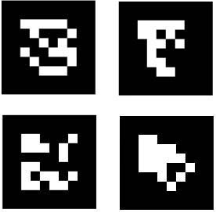}
        \caption{ARTag}
        \label{fig:artag}
    \end{subfigure} 
    \begin{subfigure}[t]{0.0925\textwidth}
        \includegraphics[width=\textwidth]{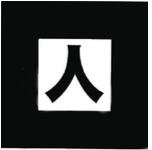}
        \caption{ARToolKit}
        \label{fig:artoolkit}
    \end{subfigure}
    \begin{subfigure}[t]{0.1025\textwidth}
        \includegraphics[width=\textwidth]{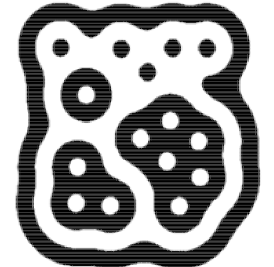}
        \caption{reacTIVision}
        \label{fig:reactivision}
    \end{subfigure}
    \begin{subfigure}[t]{0.09\textwidth}
        \includegraphics[width=\textwidth]{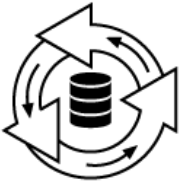}
        \caption{d-touch}
        \label{fig:dtouch}
    \end{subfigure} 
    \begin{subfigure}[t]{0.0875\textwidth}
        \includegraphics[width=\textwidth]{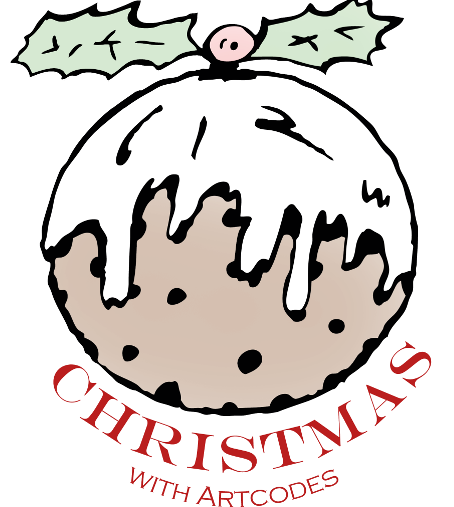}
        \caption{Artcode}
        \label{fig:artcode}
    \end{subfigure} 
    \caption{Visual marker examples.}
    \label{fig:markerExamples} 
\end{figure}

\begin{figure*}[h]
    \centering  
    \begin{subfigure}[b]{\textwidth}
        \includegraphics[width=\textwidth]{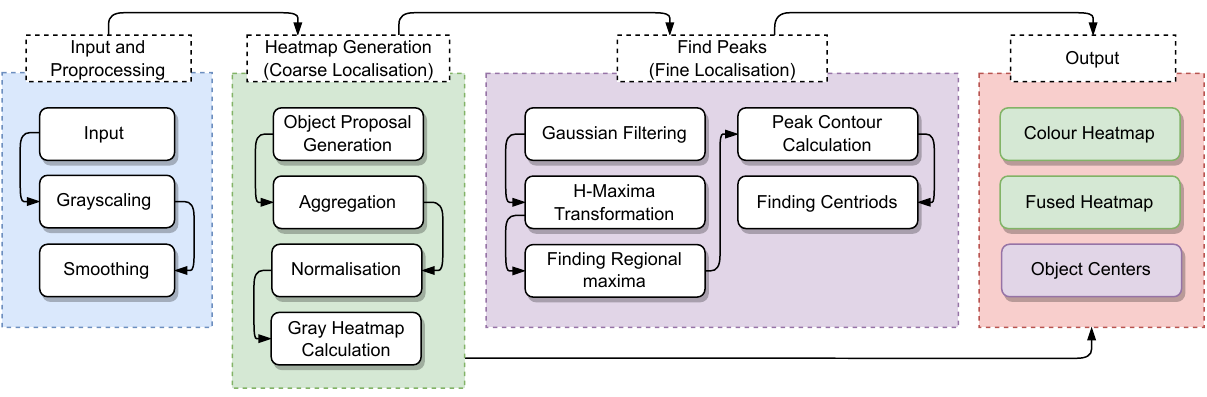}
    \end{subfigure} 
    \caption{Illustration of the framework of the {\sc VisionGuide} approach.}
    \label{fig:ArtcodeGuideFramework}
\end{figure*}

\subsection{The Artcode Approach}
\label{subsec:artcode}
Artcodes\footnote{\url{https://www.artcodes.co.uk/}} were developed as human-designable topological visual markers. 
They draw from the d-touch system \cite{costanza2009designable}.
By incorporating additional drawing constraints and aesthetic enhancements, 
Artcodes offer more visually appealing and interactive patterns compairied to d-touch \cite{meese2013codes}.
Examples of d-touch and Artcode markers can be seen in \autoref{fig:dtouch} and \autoref{fig:artcode}.
An Artcode typically comprises two components: a recognizable foreground (such as the strawberry in \autoref{fig:artcode}) and an image-based background (like the text in the same figure). 
While the foreground is designed for machine readability, the background can be tailored for human consumption.
Artcodes present an opportunity to create attractive and interactive motifs that adorn everyday objects without compromising their aesthetic appeal, unlike QR codes.

Due to their unobtrusive and non-obvious properties, Artcodes are often not immediately noticeable, requiring close inspection for discovery when visual cues are absent.
Detecting Artcodes based on their general visual features and identifying their likely locations using a heatmap is, thus, a meaningful approach. 
Interested readers are encouraged to explore the literature for more information about Artcodes, including their design, detection, and identification \cite{meese2013codes, costanza2003d, xu2017recognizing, xu2019artcode, MRArtcodesJSS2021}.

\subsection{Visual Markers}
\label{sec:visual_markers}
Various visual markers, both human-readable and machine-readable, have been proposed (see examples in \autoref{fig:markerExamples}), with two of the most well-known being barcodes \cite{woodland1952classifying} and QR codes \cite{qrcode2015}.
While barcodes are primarily used in the retail sector, QR codes have become ubiquitous \cite{meese2013codes}.  
Barcodes and QR codes were designed to be reliably read by machines without errors occurring when scanned.
However, this reliability comes at a cost of limited aesthetics: neither are visually meaningful to humans, and it can be challenging to distinguish different codes through visual inspection alone.

Many other visual marker systems share similarities with barcodes and QR codes, often encoding information within a matrix of black and white dots, and incorporating error detection and correction mechanisms. 
An example of such a marker system is the Data Matrix \cite{datamatrix2006}.
While effective for data encoding, these markers were not designed for camera-pose estimation and calibration, and are therefore unsuitable as fiducials in AR systems.
Fiducial systems, such as 
ARTag (\autoref{fig:artag}) \cite{fiala2005artag},
ARToolkit (\autoref{fig:artoolkit}) \cite{kato1999marker}, and
reacTIVison (\autoref{fig:reactivision}) \cite{bencina2005improved} serve the purpose of enabling estimation of relative pose between the camera and object.
ARTag markers utilise a square border for localisation, while ARToolkit markers consist of a thick square black border with a variety of patterns in the interior
---
the black outline allows for marker localisation and 
\emph{homography} calculation.
The reacTIVision markers are automatically generated and offer limited customisation options for aesthetic aspects \cite{bencina2005improved}.

Maker systems relying on geometrical features for localisation and encoding are often constrained in their visual appearance. 
In most cases, the shape (the geometry) of the markers is automatically generated, providing limited design flexibility.
In contrast, topological marker systems such as d-touch and its variant Artcodes \cite{meese2013codes, xu2019artcode} offer much more freedom in geometric form, allowing users to customise both the outline shape and interior elements.
d-touch encodes information through the topological structure of the markers, representing the
adjacency information of connected components in a region adjacency tree \cite{costanza2003region}.
This enables users to create their own readable markers that are both aesthetic and meaningful \cite{costanza2009designable}.
The Artcode approach further refines the d-touch approach, enhancing drawing rules and introducing human-meaningful embellishments and aesthetic style guidelines. 
The approach grants creative freedom to produce visually appealing \emph{and} machine-readable markers that are meaningful to humans and resemble free-form images. 
We thus adopt this marker system for creating the connections with the metaverse, and studying their coarse-to-fine localisation.

In addition to geometry-based visual markers, conventional image recognition technologies such as Blippar \cite{blippar2021} and Google Lens \cite{google2021lens} embed data into a wider array of images.
However, these methods, often utilising neural networks and vector matching, can be challenging to explain to non-technical users. 
Recent advancements include systems such as E2ETag \cite{brennan2021e2etag}, Artcoder \cite{su2021artcoder}, and DeepFormableTag \cite{yaldiz2021deepformabletag}.

\subsection{Connections with Metaverse}
The term ``metaverse'' was coined by Neal Stephenson in his 1992 science-fiction novel \emph{Snow Crash} \cite{stephenson2003snow}, portraying a 3D virtual world where individuals interact via avatars\cite{benford2021metaverse}. 
Despite three decades of development, the metaverse still lacks a universally accepted definition \cite{nevelsteen2018virtual, duan2021metaverse, benford2021metaverse}. 
Various frameworks and characteristics of the metaverse have been studied in the literature.
For example, Benford et al. \cite{benford2021metaverse} describe five metaverse elements: 
a virtual world,
a virtual reality, 
persistence,
connection to the real world, and 
interaction with other people. 
In contrast, Duan et al. \cite{duan2021metaverse} proposed a three-layer metaverse development architecture, representing the physical world, interaction, and the virtual world. 
Despite the lack of consensus on definition, all these frameworks share a common element---the connection between the virtual and physical worlds. 
In this study, we employ the Artcode approach to connect physical worlds (everyday environments) with virtual worlds (the metaverse).

\section{The Coarse-to-fine Approach}
\label{sec:approach}
The Artcode approach enables the creation of interactive surface patterns that can augment objects in everyday environments.
However, as described in \autoref{subsec:artcode}, the variability in the shape and appearance of Artcodes presents challenges in detecting their presence.
Consequently, the subsequent decoding of Artcodes to trigger the associated digital materials cannot be achieved, resulting in a failure to connect with the metaverse.
To address this issue, we propose a vision-based coarse-to-fine approach---{\sc VisionGuide}---to progressively localise Artcodes. 
This approach guides individuals in identifying potential locations using a heatmap that indicates where Artcodes are likely to be found.

As illustrated in \autoref{fig:ArtcodeGuideFramework}, {\sc VisionGuide} comprises two primary steps: heatmap generation (coarse localisation) and peak finding (fine localisation). 
There are two additional steps: input and preprocessing, and output).
We describes these steps in detail in the following sections.

\subsection{Input and Preprocessing}
\label{subsec:input}
The first step of {\sc VisionGuide} is preprocessing, which includes grayscaling and smoothing. 
Artcodes primarily rely on their topological features for detection, as geometrical and visual features are ineffective in this regard \cite{xu2017recognizing}. 
The recognisable structure of an Artcode comprises connected regions \cite{xu2017recognizing}. 
To enhance detection, the input image is first converted into grayscale and then smoothed using bilateral filtering \cite{tomasi1998bilateral}. 
Bilateral filtering, a non-iterative approach for image smoothing, effectively preserves edge details. 
Widely used in computer vision and graphics tasks \cite{porikli2008constant}, bilateral filters differ from conventional spatial domain filters such as a Gaussian filter (GF) by considering both spatial distance and intensity disparity from the central pixel when assigning coefficients. 
Consequently, different pixels in an image may undergo different bilateral filters.

This preprocessing step is crucial for the performance of the proposed approach. 
The impact of bilateral filtering on the {\sc VisionGuide}'s performance is experimentally studied in \autoref{sec:experiments}.

\begin{figure}[!t]
    \centering  
    \begin{subfigure}[b]{0.20\textwidth}
        \includegraphics[width=\textwidth]{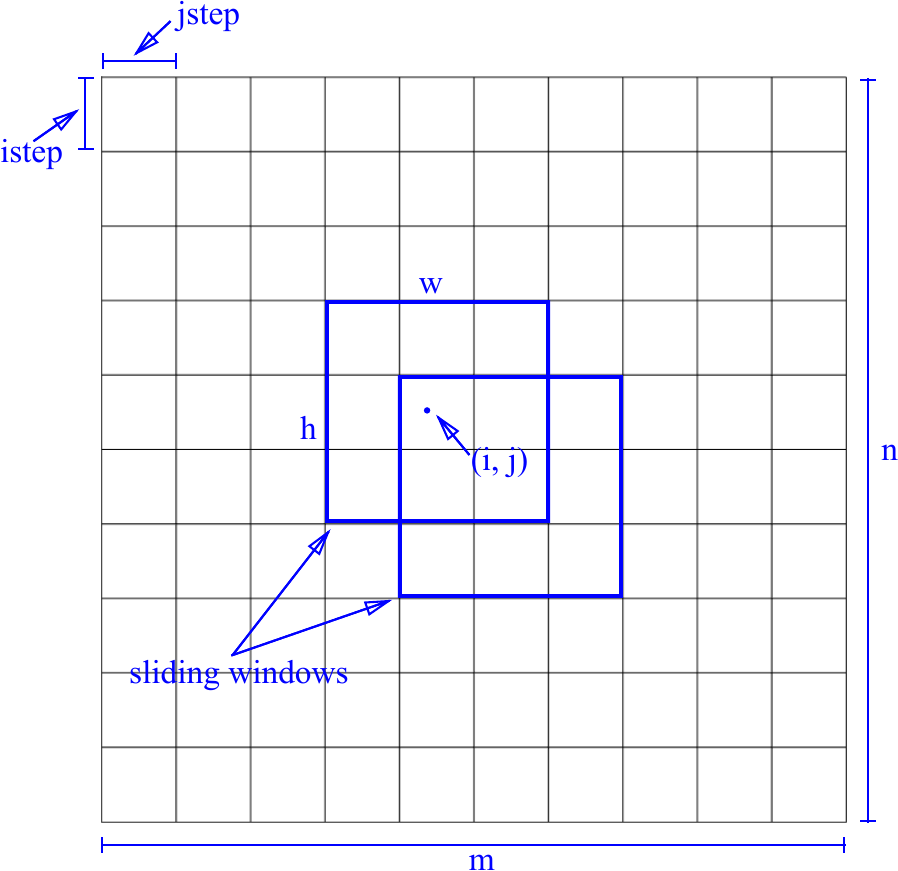}
    \end{subfigure} 
    \caption{Illustration of the sliding window and the moving steps.}
    \label{fig:slidingWindowIllustration}
\end{figure}

\begin{figure}
    \centering  
    \begin{subfigure}[b]{0.475\textwidth}
        \includegraphics[width=\textwidth]{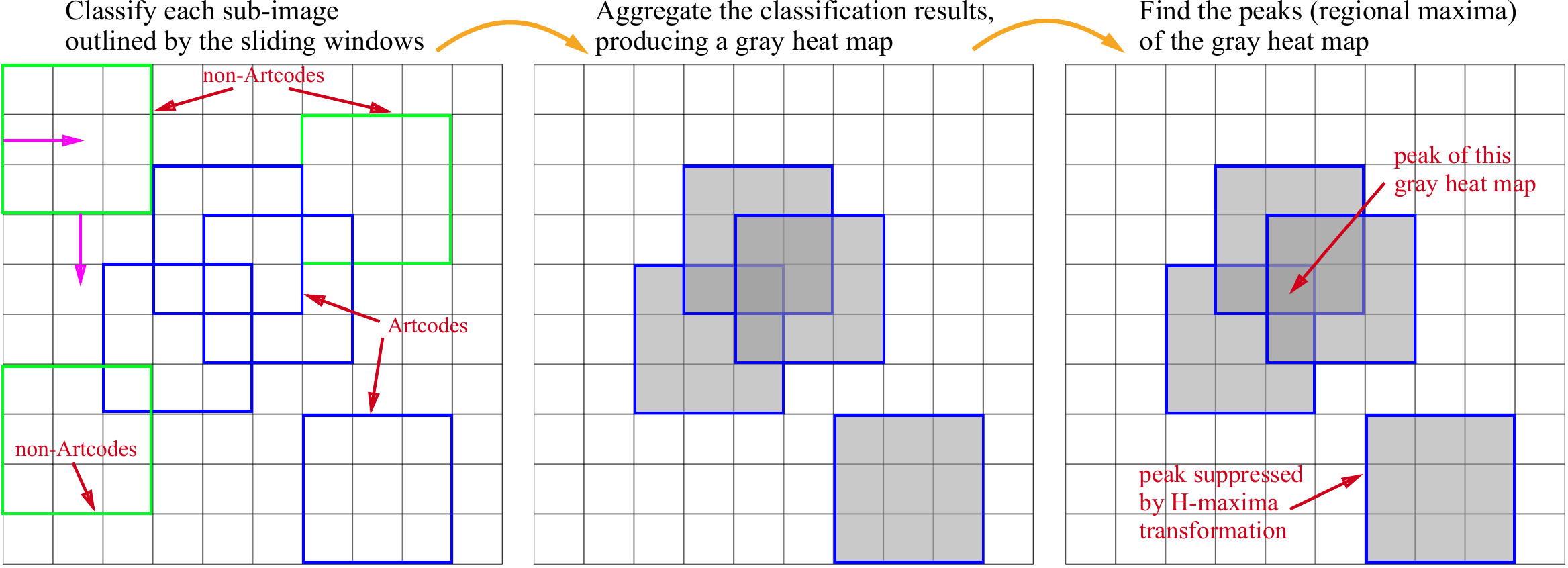}
    \end{subfigure} 
    \caption{
        Illustration of the heatmap calculation process and the subsequent peak-finding procedure. 
        The two pink arrows in the left figure indicate the moving direction of the sliding window, which first moves horizontally and then vertically. 
        The blue and green boxes are sliding window examples classified by the trained classifier as containing either {\it non-Artcode} or {\it Artcode}. 
        In the middle figure, the aggregation results of the left figure are shown, where the darkness level represents the likelihood of each area containing Artcodes---darker areas indicate a higher likelihood of an Artcode presence. 
        The right figure illustrates the procedure for finding regional maxima.
    }
    \label{fig:calcHeatmapIllustration}
\end{figure}

\subsection{Heatmap Generation}
\label{subsec:heatMapCalculation}
\begin{algorithm}[!t]
\caption{Heatmap calculation algorithm.}
\label{algo:calcHeatmap}
\begin{algorithmic}[1]
    \Procedure{calcHeatMap}{$im, slidewin, istep, jstep, h$}
    \State $ [n, m] \gets size(im)$ 
    \State $heatmap \gets zeros(n, m)$
    \State $im \gets \textproc{bilateralFilter}(im)$  \label{line:bilateralFilter}
    
    \For{$i \gets 1:istep:n$}
        \For{$j \gets 1:jstep:m$}
            \State $subim \gets \textproc{imCrop}(i, j, slidewin)$  \label{line:imcrop} 
            \State $features \gets \textproc{calcSOHFeature}(subim)$ \label{line:sohFeature} 
            \State $[class, score] \gets h(features)$ \label{line:classify} \Comment{$h$ is a trained classifier.}
            \If{$class = 1$}
                \State $h \gets slidewin[1]$ 
                \State $w \gets slidewin[2]$
                \State $heatmap(i:i+h, j:j+w) \gets heatmap(i:i+h, j:j+w)+score$  \label{line:aggregate}
            \Else
                \State $continue$ 
            \EndIf
        \EndFor
    \EndFor
    \State \textbf{return} $heatmap$ 
    \EndProcedure
\end{algorithmic}
\end{algorithm}

\begin{figure}[t]
        \centering  
        \begin{subfigure}[t]{0.1175\textwidth}
            \includegraphics[width=\textwidth]{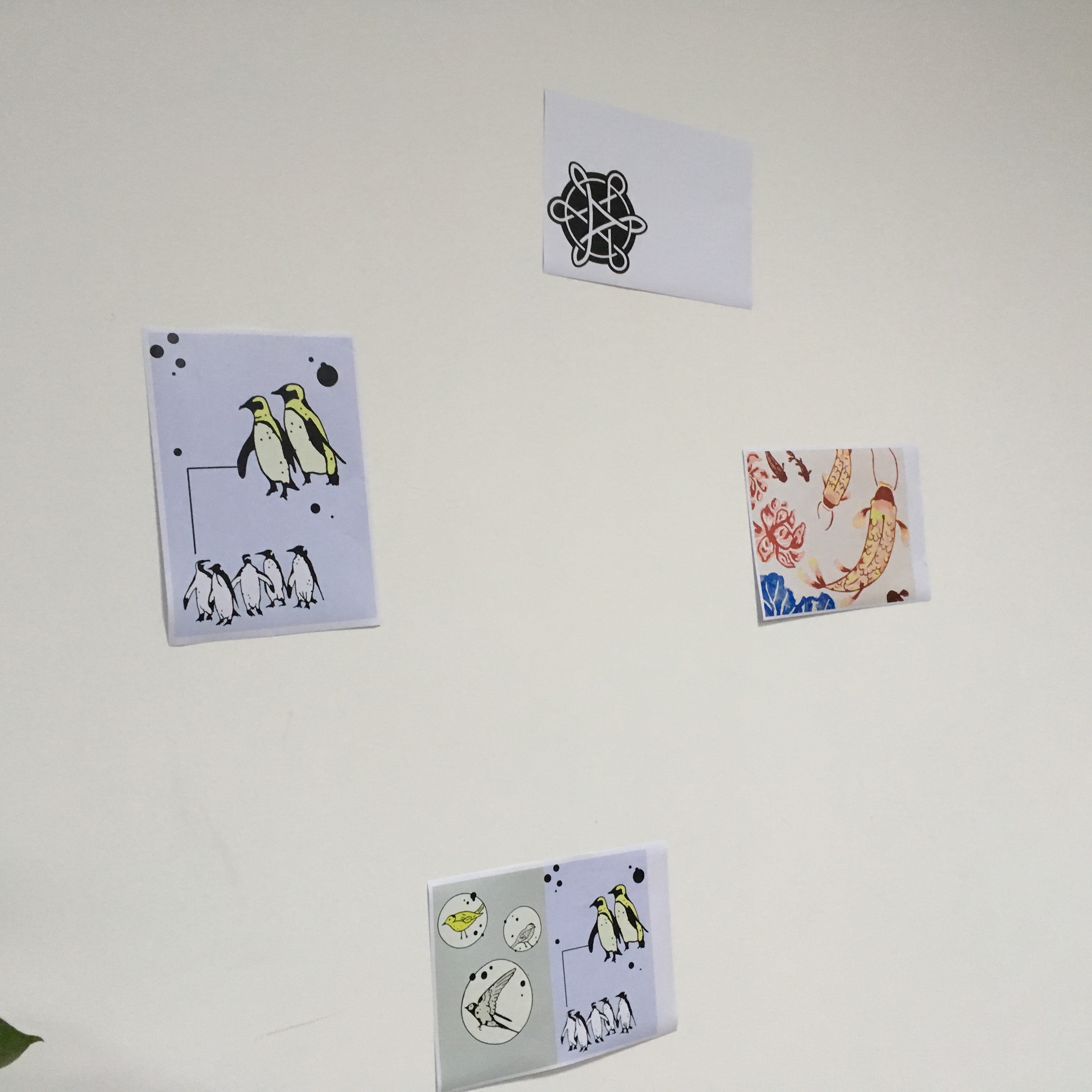}
            \caption{Input}
            \label{fig:heatmap-input}
        \end{subfigure} 
        \begin{subfigure}[t]{0.118\textwidth}
            \includegraphics[width=\textwidth]{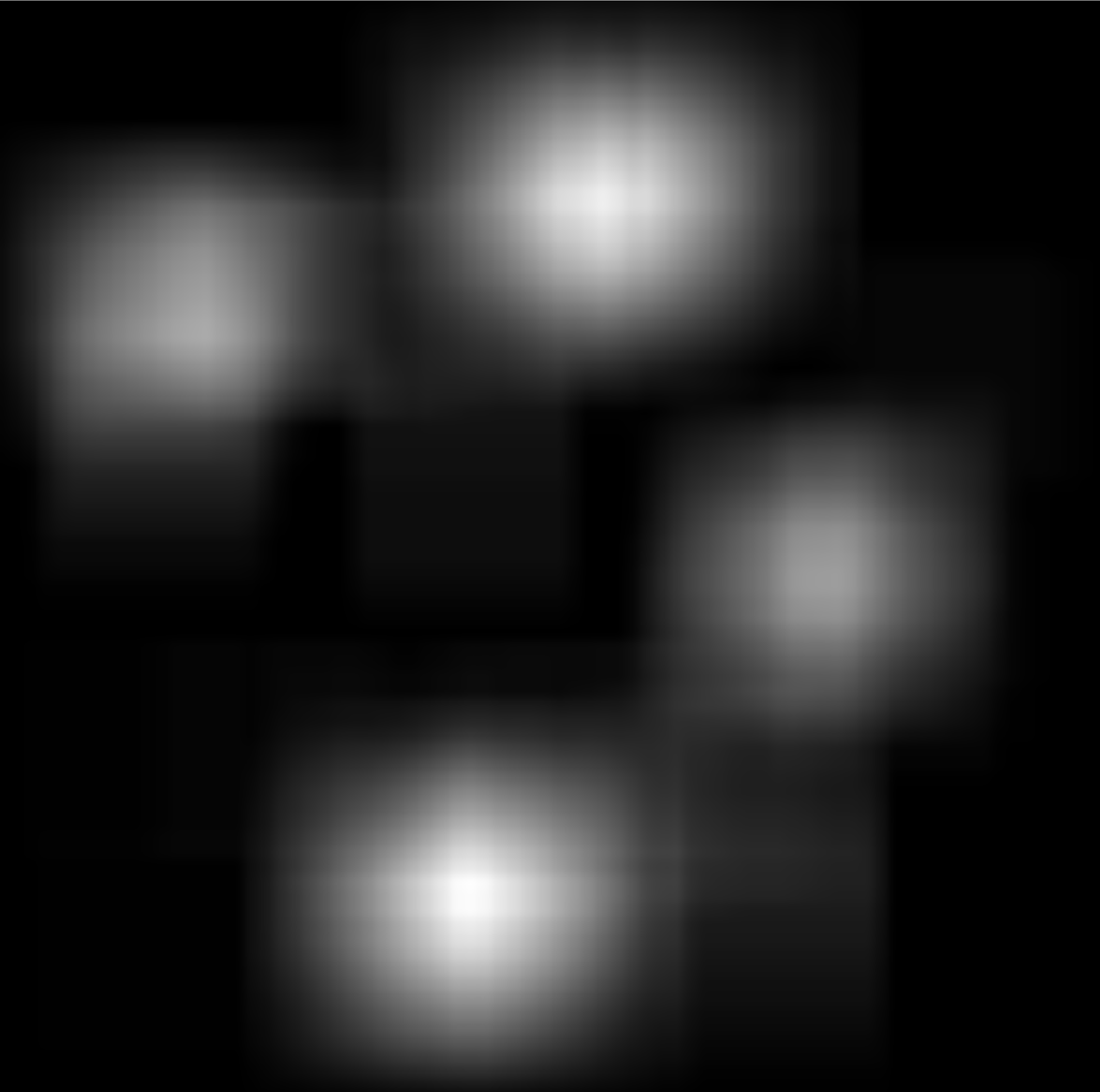}
            \caption{Grayscale}
            \label{fig:heatmap-gray}
        \end{subfigure} 
        \begin{subfigure}[t]{0.1175\textwidth}
            \includegraphics[width=\textwidth]{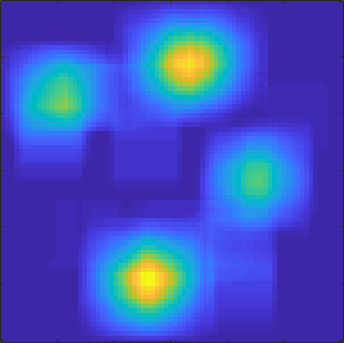}
            \caption{Colour}
            \label{fig:heatmap-color}
        \end{subfigure}
        \begin{subfigure}[t]{0.1175\textwidth}
            \includegraphics[width=\textwidth]{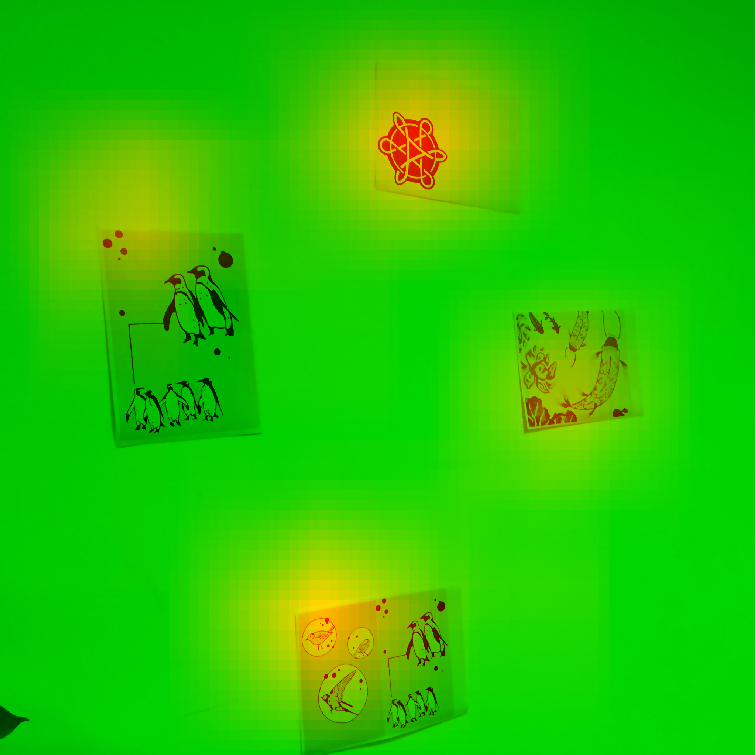}
            \caption{Fused}
            \label{fig:heatmap-fused}
        \end{subfigure}
        \caption{Input image and its corresponding heatmaps. 
            (\subref{fig:heatmap-gray}) and (\subref{fig:heatmap-color}) are heatmaps generated by \autoref{algo:calcHeatmap}, plotted in grayscale and colour space respectively. 
            In these heatmaps, the brighter (higher energy) the area, the higher the likelihood of containing Artcodes.
            (\subref{fig:heatmap-fused}) is the fused result of the gray heatmap with the input image.}
         \label{fig:heatmap-example}
\end{figure}

The smoothed grayscale image is then fed into the next step for heatmap generation.
A heatmap is a two-dimensional representation of data, with values visualised through colours. 
It offers a visual summary of the presence of the Artcodes in the image. 
In this study, a two-dimensional heatmap with the same size as the input image was used. 
The value (colour) of the heatmap at point $(i, j)$ indicates the likelihood of that point being part of an Artcode.
Refer to \autoref{fig:heatmap-example} for a visualisation of an input image and its corresponding heatmaps.

The detailed algorithm for calculating the heatmap is presented in \autoref{algo:calcHeatmap}. 
The main idea behind heatmap calculation involves sliding a sub-window (sliding window) over the input image. 
The sub-images contained in the sliding window are then classified using a trained classifier (\lineautorefname\ref{line:imcrop} in \autoref{algo:calcHeatmap}). 
Subsequently, the classification results of all sub-images (Artcode proposals) are aggregated to generate a heatmap (\lineautorefname\ref{line:aggregate}). 
The process of heatmap calculation is illustrated in \autoref{fig:calcHeatmapIllustration}. 
The sliding window moves horizontally from left to right, row by row, as depicted by the pink arrows in the left figure of \autoref{fig:calcHeatmapIllustration}. 
This movement generates numerous sliding windows, determined by parameters: $istep$, $jstep$, $slidewin = [w, h]$. 
The total number of sliding windows for an $n \times m$ image is: $\ceil*{\frac{(n-h)}{ istep}} \times \ceil*{\frac{(m-w)}{ jstep}} $, where $\ceil{x}$ returns the nearest integer that is greater than or equal to $x$.

The sub-images are classified as either {\it Artcode} or {\it non-Artcode} by a trained classifier. 
Each classification returns a {\it class} and its corresponding {\it score}, representing the likelihood of the image containing Artcodes. 
If a sub-image is classified as containing Artcodes (class = 1), the heatmap is updated accordingly; the values of points within the region covered by this sub-image are incremented by the prediction score. 
This step involves aggregating the classification results. 
As illustrated in \autoref{fig:calcHeatmapIllustration}, the sub-images outlined by blue boxes are predicted to contain Artcodes (or parts of Artcodes), while those outlined by green boxes are predicted to contain no Artcodes. 
The scores associated with the blue boxes are used to update the heatmap, resulting in the middle figure in \autoref{fig:calcHeatmapIllustration}.
In this heatmap representation, the central darkest area (overlapped by three sub-images) attains the highest score.
The adjacent areas covered by two sub-images appear darker than their boundaries, which are covered by only one sub-image. 
The bottom-right area is also brighter than the central region due to its lower score.

The entire process results in a {\it score} or {\it likelihood} matrix, where the value at point $(x, y)$ represents the total likelihood of this point being contained by an Artcode. 
If all the sliding windows containing this point are Artcodes, then the value of the heatmap at point $(i, j)$ obtains the maximum score, given by: 
\begin{equation}\label{eq:accumulatedScore}
    \max\limits_{\substack{1 \leq i \leq n \\ 1 \leq j \leq m}} heatmap(i, j) = \sum_{k = 1}^N S_k
\end{equation}
where $N$ is $
    \Big( \floor*{\frac{\min(j, m - w)}{jstep}} - \ceil*{\frac{j-w}{jstep}} + 1 \Big) \times \Big( \floor*{\frac{\min(i, n-h)}{istep}} - \ceil*{\frac{i-h}{istep}} + 1 \Big) $
and $S_k$ is the $score$ of sliding window $k$ classified as containing Artcodes. 
The score is zero if the sub-image is predicted to be non-Artcode. 
The resulting matrix is the output heatmap, denoted by $heatmap$, which can be normalised and plotted through different colour maps. 
This heatmap represents a coarse localisation of Artcodes, indicating the possible presence of Artcodes from a relatively greater interaction distance. 
Subsequently, this heatmap is utilised in the next stage for fine localisation, where the peaks corresponding to the highest energy spots are identified.

\subsection{Finding Peaks}\label{subsec:findingPeaks}
\begin{figure}[t]
    \centering
    \begin{subfigure}[t]{0.155\textwidth}
        \includegraphics[width=\textwidth]{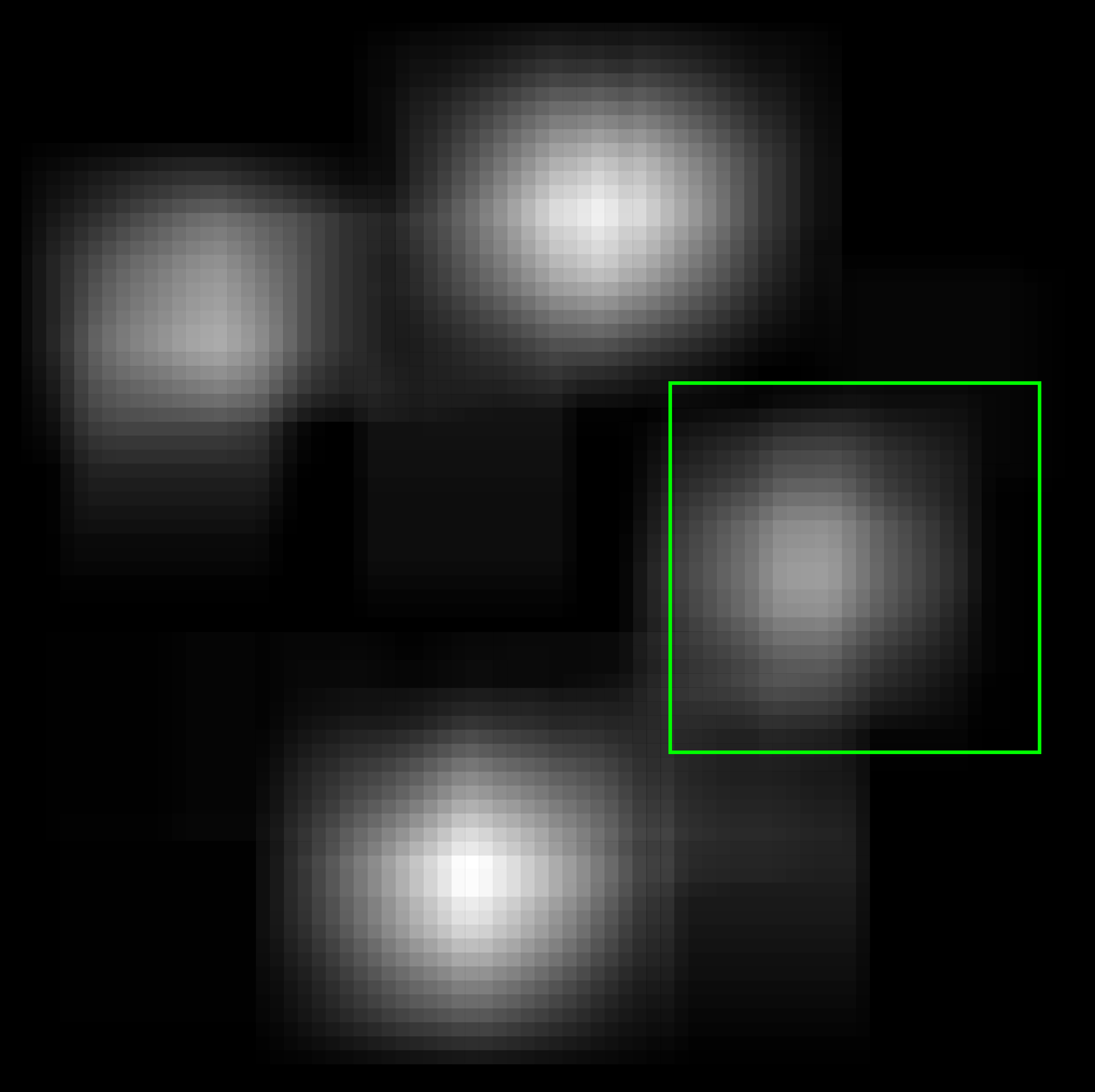}
        \caption{Original}
        \label{fig:heatmap-original}
    \end{subfigure}   
    \begin{subfigure}[t]{0.155\textwidth}
        \includegraphics[width=\textwidth]{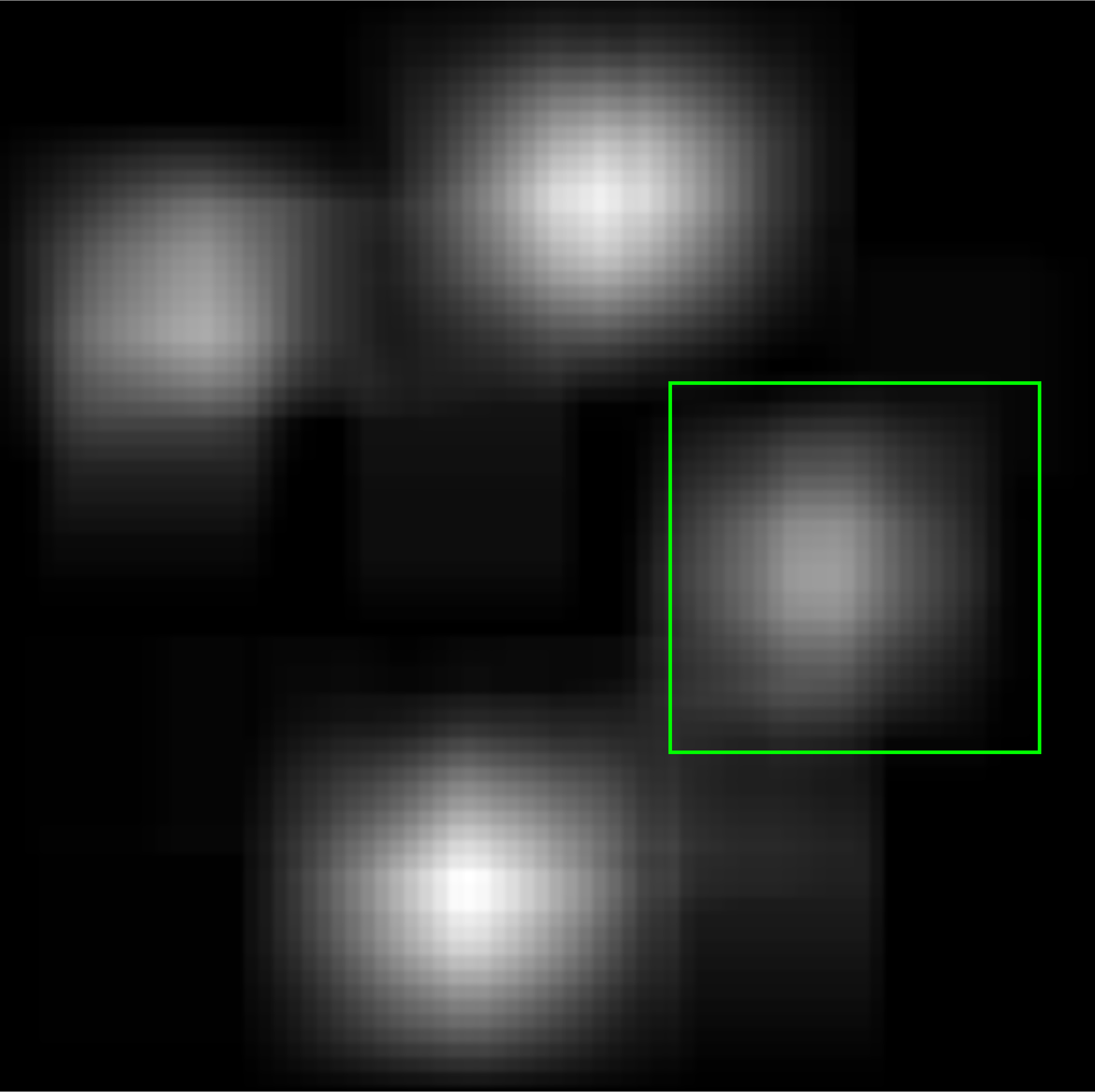}
        \caption{GF ($\sigma = 10$)}
        \label{fig:heatmap-s10}
    \end{subfigure} 
    \begin{subfigure}[t]{0.155\textwidth}
        \includegraphics[width=\textwidth]{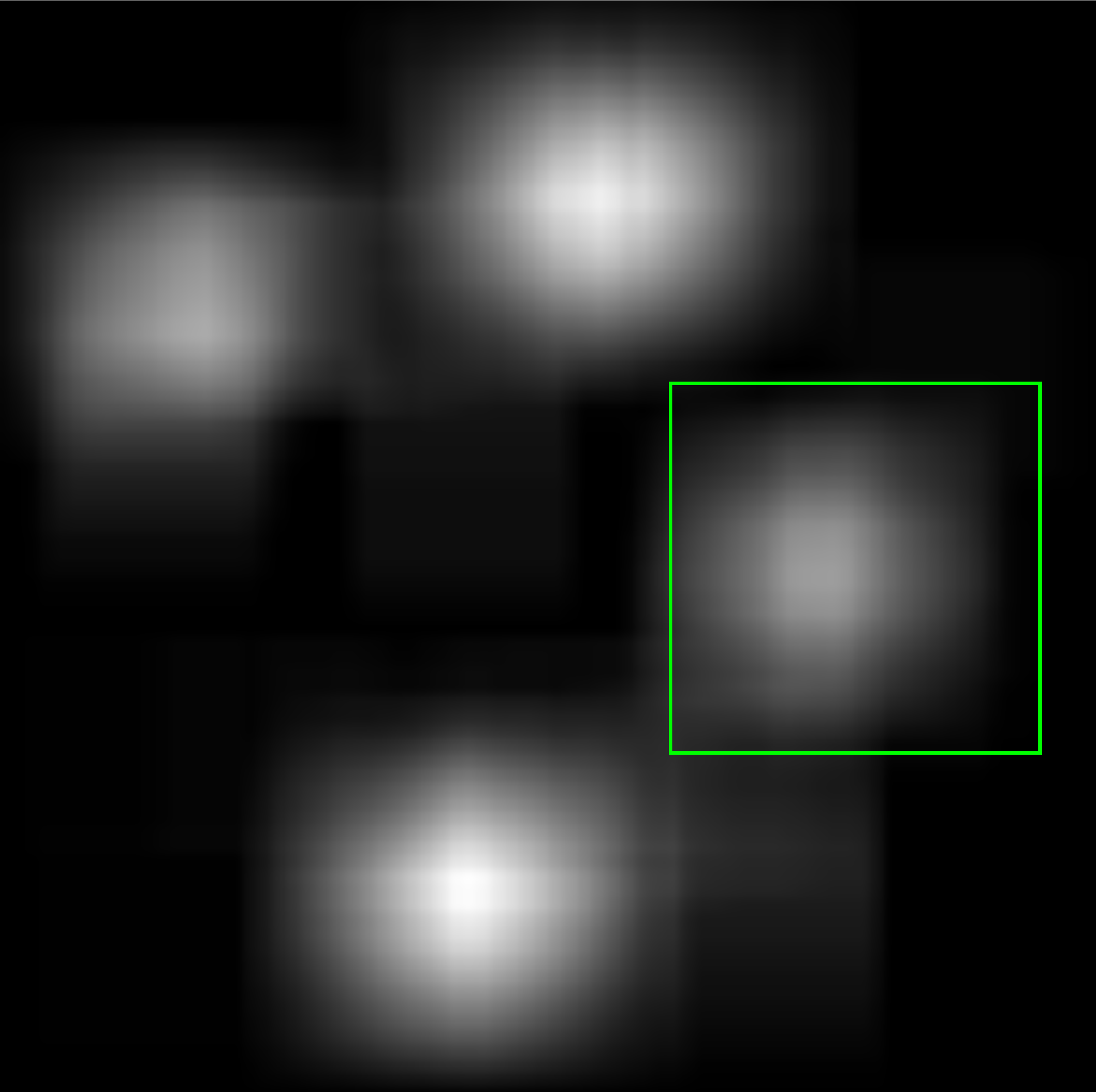}
        \caption{GF ($\sigma = 20$)}
        \label{fig:heatmap-s20}
    \end{subfigure} 
    \begin{subfigure}[t]{0.155\textwidth}
        \includegraphics[width=\textwidth]{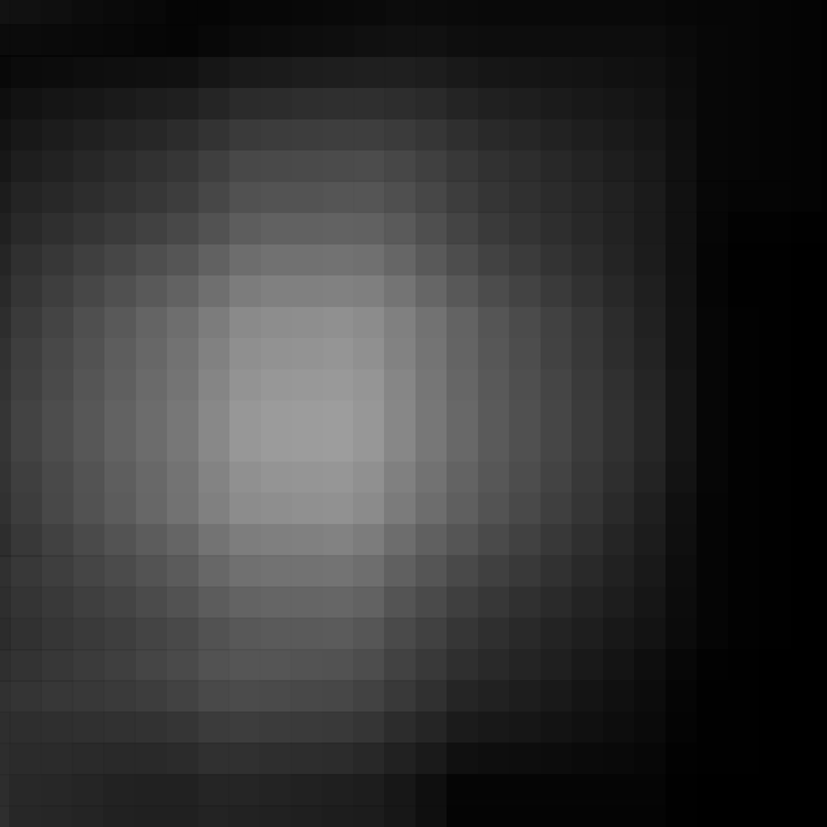}
        \caption{Cropped}
        \label{fig:heatmap-original-crop}
    \end{subfigure}   
     \begin{subfigure}[t]{0.155\textwidth}
        \includegraphics[width=\textwidth]{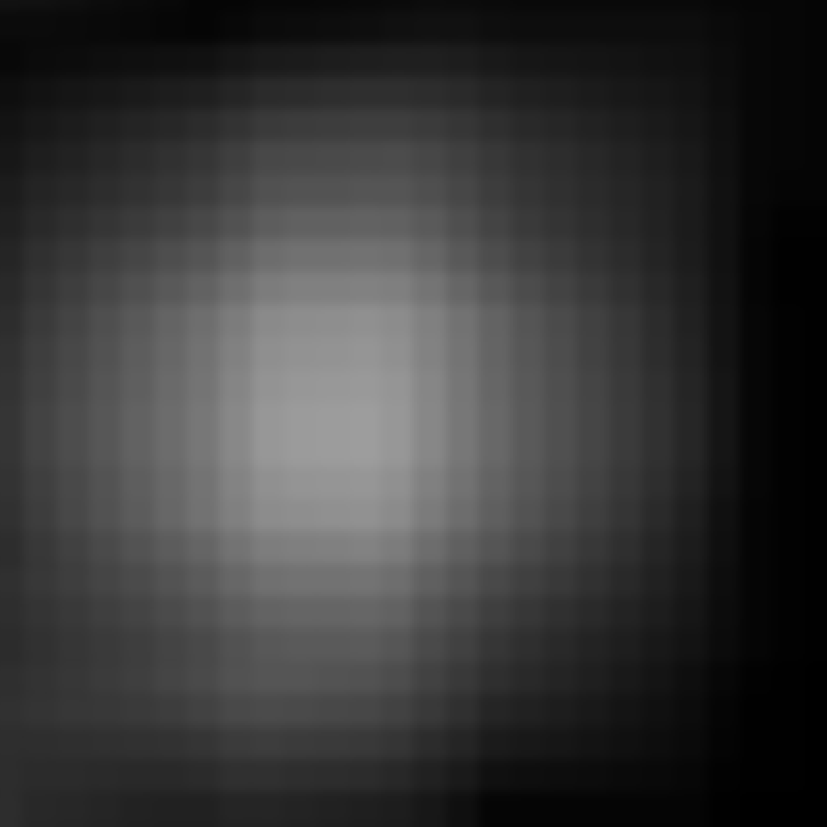}
        \caption{GF ($\sigma = 10$)}
        \label{fig:heatmap-s10-crop}
    \end{subfigure} 
    \begin{subfigure}[t]{0.155\textwidth}
        \includegraphics[width=\textwidth]{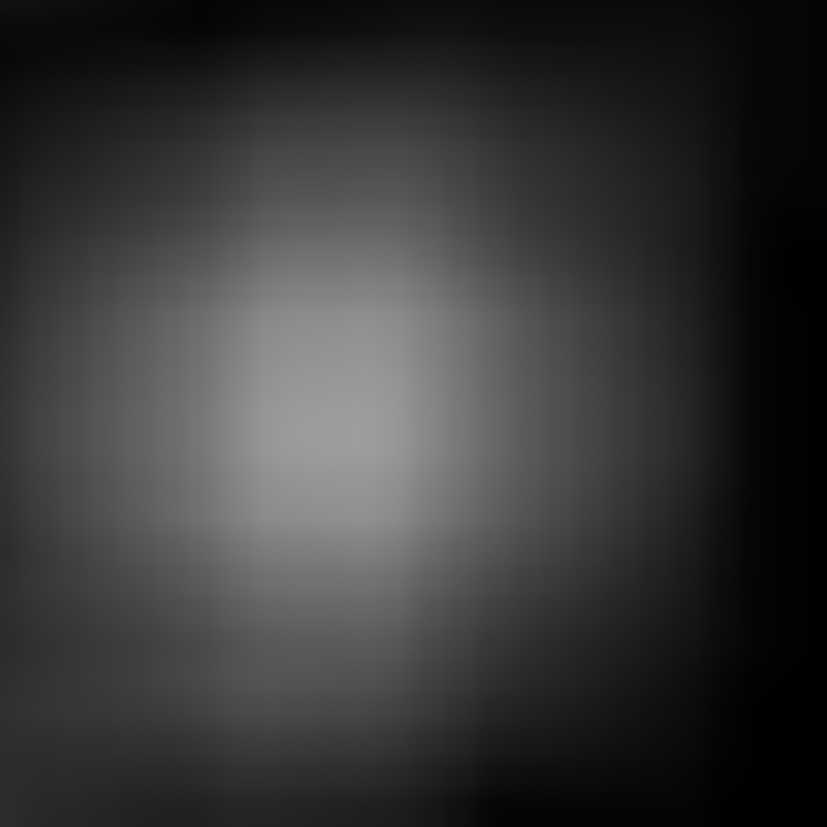}
        \caption{GF ($\sigma = 20$)}
        \label{fig:heatmap-s20-crop}
    \end{subfigure} 
    \caption{
        The original heatmap generated by \autoref{algo:calcHeatmap} with input arguments: $slidewin = [800, 800], istep = jstep = 40, h = 10$) and its smoothed ones using Gaussian filter with $\sigma=10$ and $\sigma=20$. 
        For better illustration, the bottom images are cropped from the patches annotated by the green rectangles in the top. 
    }
    \label{fig:gaussFilteredHeatmap}
\end{figure}

When the heatmap matrix is ready, it is first normalised to the range of $[0, 255]$ for improved computation and visualisation. 
Then, the {\sc findPeaks} procedure is empolyed to find the peaks, representing Artcode centres, within the heatmap.
This procedure consists of four main steps: Gaussian filtering, H-maxima transformation (HMT), regional maxima finding, and centroids finding.

\subsubsection{Gaussian Filtering}
\label{par:gaussFiltering}
To accelerate the computation, the movement of the sliding window is not {\it pixel-wise} but {\it step-wise}, with $istep$ and $jstep$ representing the horizontal (row) and vertical (column) step lengths, respectively. 
As a result, the heatmap generated is not smooth or visually continuous, but rather stepped.  
To mitigate this effect, Gaussian filtering is applied to the heatmap to achieve smoothing.

A Gaussian filter is characterised by its response being a Gaussian function, typically used for smoothing or blurring images. 
The Gaussian function is given by: 
\begin{equation}\label{eq:2dgauss}
    g(x, y) = \frac{1}{2\pi\sigma^2} \exp\bigg\{-\frac{x^2 + y^2}{2\sigma^2}\bigg\}
\end{equation}

Mathematically, applying a Gaussian filter to an image involves convolving the image with a Gaussian function. 
This spatial domain filter averages image values using weights that decrease as the distance from the centre increases. 
The convolution operation of a Gaussian filter applied to an image $\vec{f}(\vec{x})$ produces an output image as follows: 
\begin{equation}\label{eq:gaussfilterEquation}
	\vec{h}(\vec{x}) = \frac{1}{k(\vec{x})} \int_{-\infty}^{\infty} \int_{-\infty}^{\infty} \vec{f}(\xi)g(\xi, \vec{x})d\xi
\end{equation}
with the normalisation term: 
\begin{equation}
    k(\vec{x}) = \int_{-\infty}^{\infty} \int_{-\infty}^{\infty} g(\xi, \vec{x})d\xi
\end{equation}
where $g$ is the two-dimensional Gaussian function described in \autoref{eq:2dgauss}, and $g(\xi, \vec{x})$ is the Gaussian filter coefficient between $\vec{x}$ and its neighbourhood $\xi$.

As shown in \autoref{eq:gaussfilterEquation}, the value of pixel $\vec{x}$ is set to a weighted average of the neighbourhood $\xi$. z
The step effect could be reduced if an appropriate Gaussian kernel is picked. 
As shown in \autoref{fig:gaussFilteredHeatmap}, the original heatmap (\autoref{fig:heatmap-original}) generated by \autoref{algo:calcHeatmap} exhibits noticeable stepping. 
However, after Gaussian filtering with $\sigma = 10$, the heatmap appears smoother compared to the original, although the steps are still evident.

The size of the Gaussian filter is related to the sizes of the moving steps ($istep$ and $jstep$) of the sliding window. 
Larger moving steps require larger Gaussian filters to mitigate the step effect. 
In \autoref{sec:experiments}, we will experimentally investigate the impact of the filter size $\sigma$ and the moving step sizes on fine localisation.
Empirically, satisfactory results are achieved when the kernel size of the Gaussian filter is approximately equal to or greater than one-quarter of the moving steps.
For example, when $istep=jstep=40$, using $\sigma=10$ yields satisfactory outcomes for {\sc VisionGuide}.

\begin{figure}[t]
    \centering  
    \begin{subfigure}[b]{0.475\textwidth}
        \includegraphics[width=\textwidth]{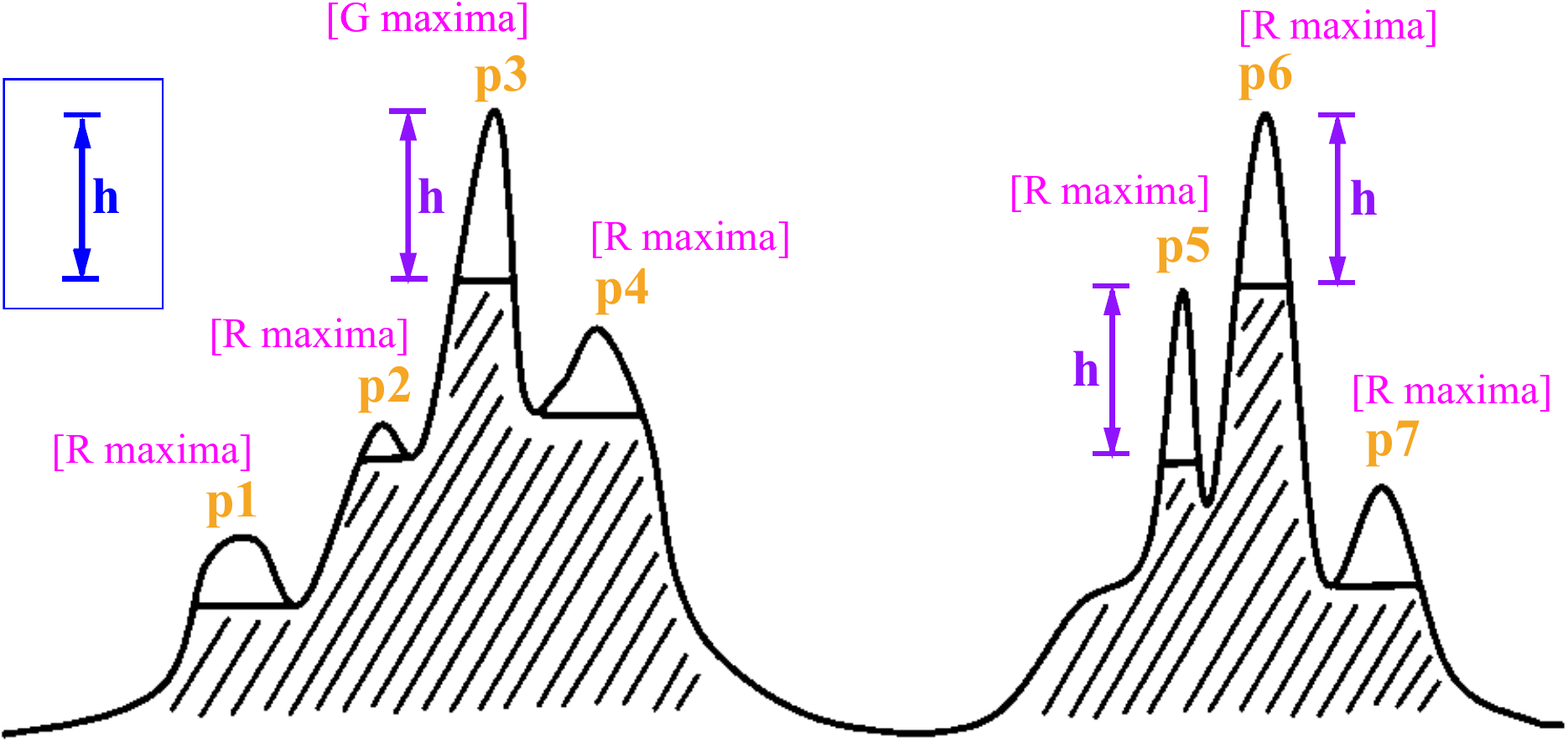}
    \end{subfigure}
    \caption{
        Illustration of the HMT and regional maxima in one dimension. 
        The peaks (regional maxima) are labelled as {\tt p1}--{\tt p7}, with each peak assigned a label to indicate its type: global maximum ({\tt G maximum}) or regional maximum ({\tt R maximum}). 
        The height $h$ in the blue box at the top right area is the {\it predefined} threshold level. 
        Peaks whose height is less than $h$ will be suppressed. 
        In this example, peaks {\tt p3}, {\tt p5}, and {\tt p6} are retained, while the others will be suppressed.
    }
    \label{fig:hmaxtransform-graphic}
\end{figure}

\subsubsection{H-Maxima Transformation}
\label{sec:hmaxTransformation}
In addition to Gaussian filtering, HMT is employed to suppress the lower peaks (maxima).
HMT \cite{soille2013morphological} is a morphological operation that retains significant maxima in a grayscale image while suppressing less significant ones based on a threshold level, $h$.
Peaks whose height is less than $h$ are suppressed by the operation. 
As illustrated in \autoref{fig:hmaxtransform-graphic}, peaks labelled {\tt p1}--{\tt p7} represent regional or local maxima (denoted {\tt R maxima}), with {\tt p3} being the global maximum ({\tt G maximum}).
In this example, only peaks {\tt p3}, {\tt p5} and {\tt p6} have heights greater than the threshold $h$, and thus, they are retained while the others ({\tt p1}, {\tt p2}, {\tt p4}, and {\tt p7}) are eliminated.

In an image, small intensity fluctuations can represents regional minima or maxima. Similarly, a Gaussian-smoothed heatmap may contain numerous less significant peaks that do not indicate the presence of Artcodes. 
These minor peaks could be caused by the background, particularly the background imagery of the Artcode itself.
Therefore, HMT is used to eliminate peaks with heights below a predefined threshold $h$, thereby reducing the number of false positives.

\subsubsection{Finding Regional Maxima}
\label{par:regionalMaxima}
After Gaussian filtering and HMT, the remaining significant regional maxima of the heatmap are much more likely to have been produced by Artcodes. 
The step first identifies the general region of the Artcode by finding the regional maxima of the heatmap. 
Regional maxima ({\tt p1}--{\tt p7} in \autoref{fig:hmaxtransform-graphic}) are connected components (the connected pixels with constant intensity values), where all external boundary pixels have lower intensity values compared with the internal pixels. 
The regions output from this step represent the central areas of the Artcodes.

\subsubsection{Finding Centroids}
Finally, the last of step of finding peaks involves calculating the centroids of these regions. 
This step is straightforward, utilising a common centroid-finding approach to determine the centres of these regions. 
These centroids are the exact locations of the Artcodes embedded in the input images and are also an output of the {\sc VisionGuide} approach.

\subsection{Output}
\label{sec:output}
The two primary outputs of {\sc VisionGuide} are the presence heatmap of Artcodes and the centroids of Artcodes, generated from the corresponding two main steps: coarse and fine localisation.
When visualising the heatmap as an independent output, it can be displayed by fusing with the input image, resulting a fused heatmap as shown in \autoref{fig:heatmap-example}.
Meanwhile, the heatmap can be fed into next step for fine localisation, outputting the exact centres of the Artcodes.

\section{Experimental Studies}
\label{sec:experiments}
\begin{figure*}[h]
    \centering
    \begin{subfigure}[t]{0.095\textwidth}
        \includegraphics[width=\textwidth]{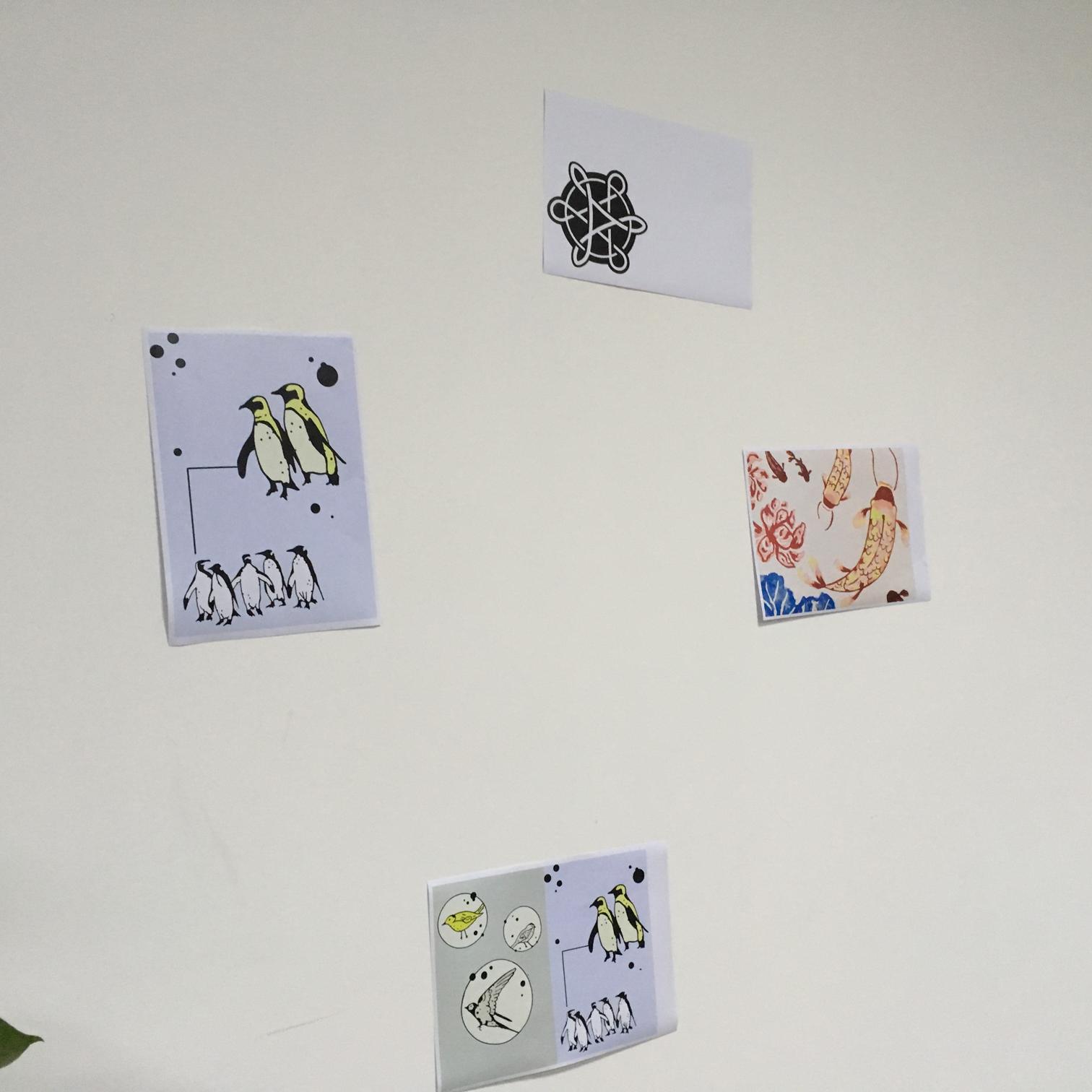}
    \end{subfigure}
    \begin{subfigure}[t]{0.095\textwidth}
        \includegraphics[width=\textwidth]{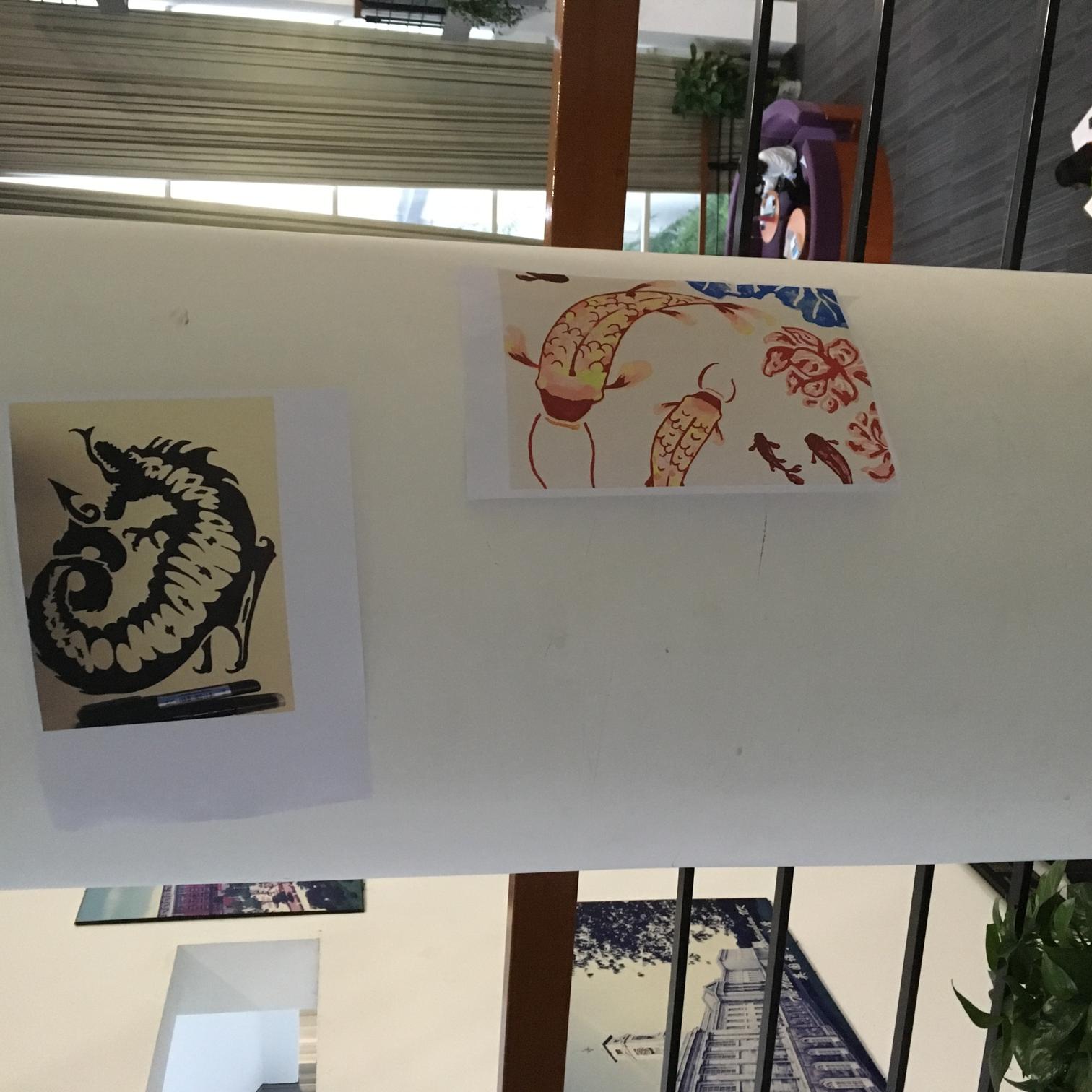}
    \end{subfigure} 
    \begin{subfigure}[t]{0.0965\textwidth}
        \includegraphics[width=\textwidth]{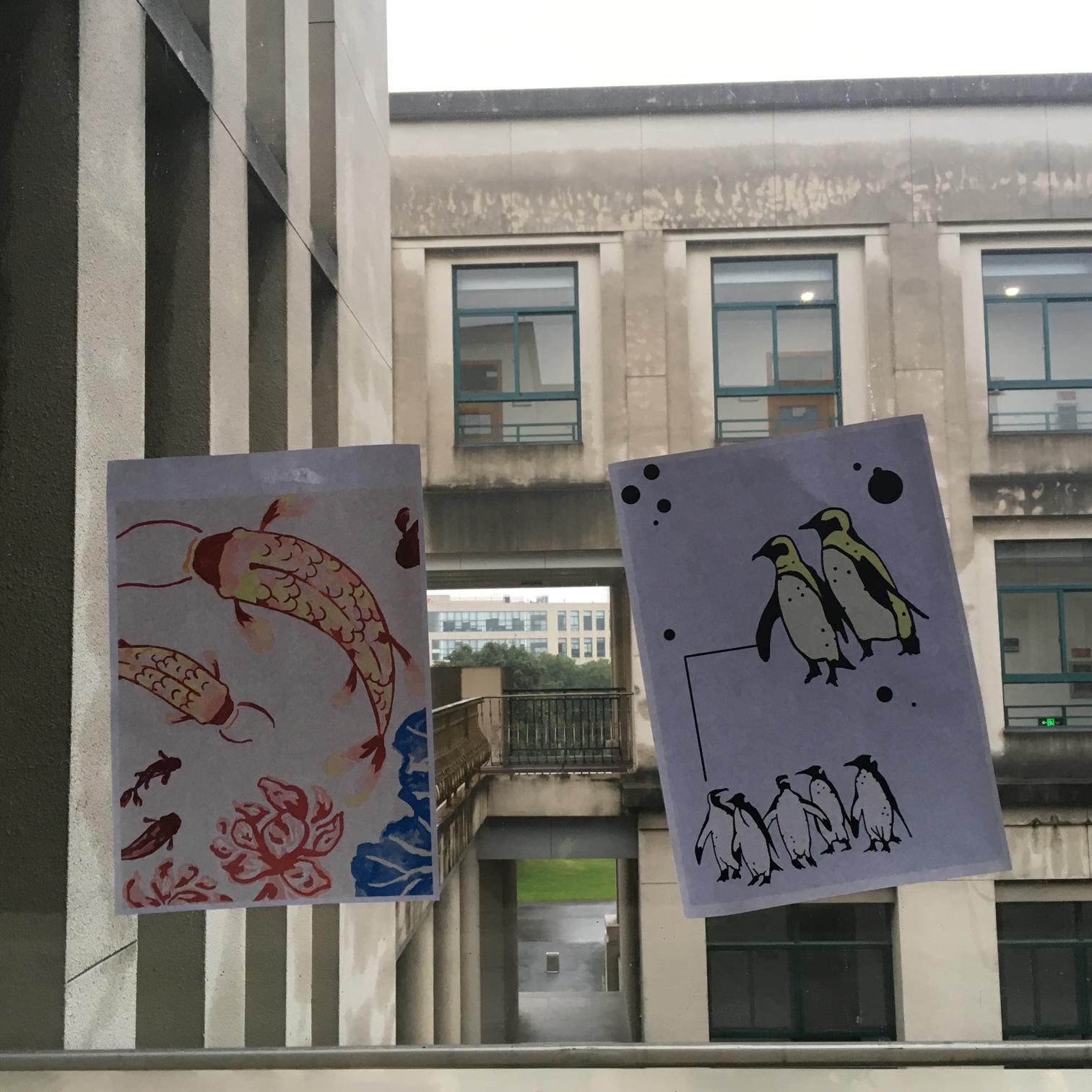}
    \end{subfigure} 
    \begin{subfigure}[t]{0.1245\textwidth}
        \includegraphics[width=\textwidth]{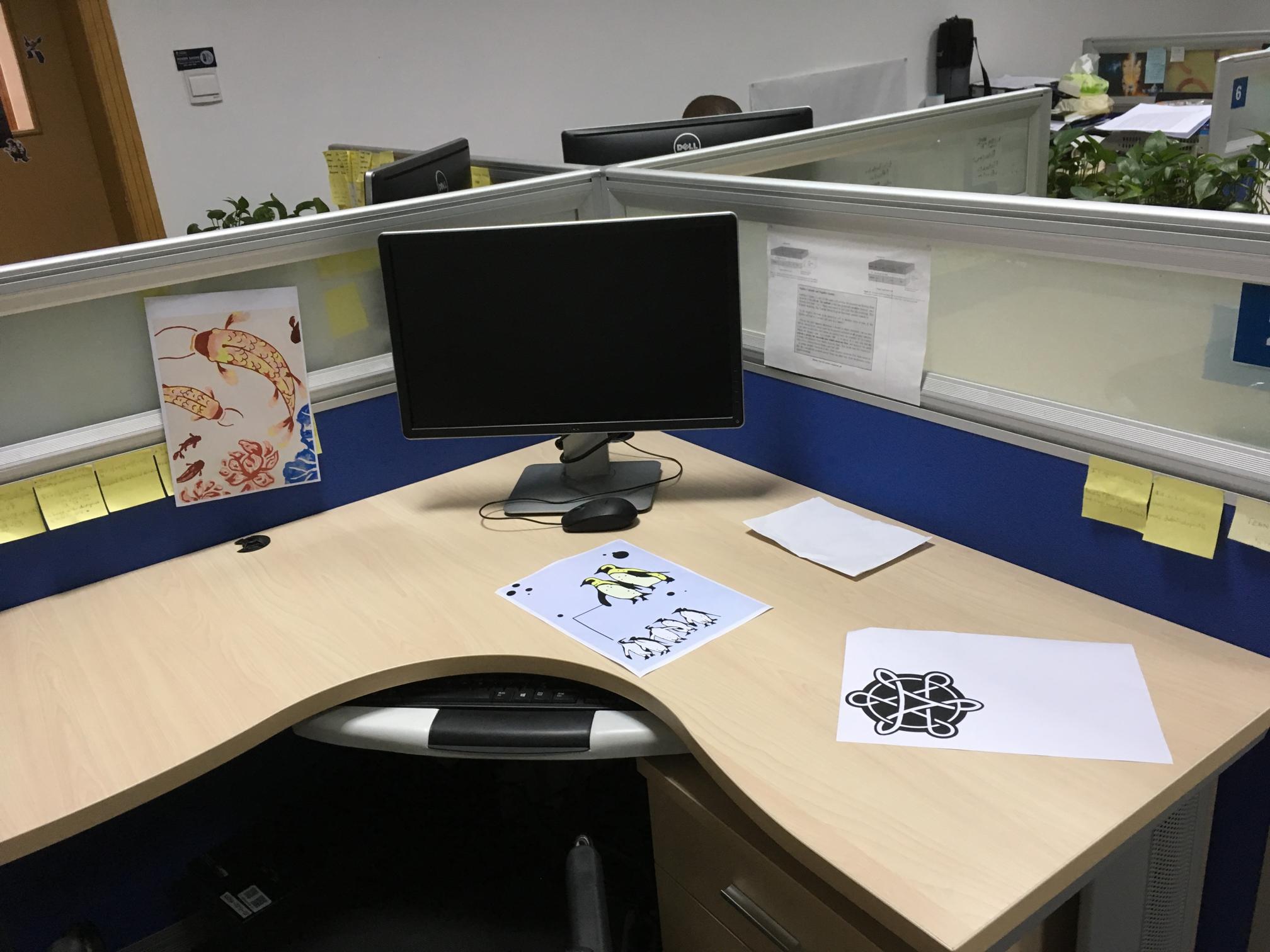}
    \end{subfigure}        
    \begin{subfigure}[t]{0.1245\textwidth}
        \includegraphics[width=\textwidth]{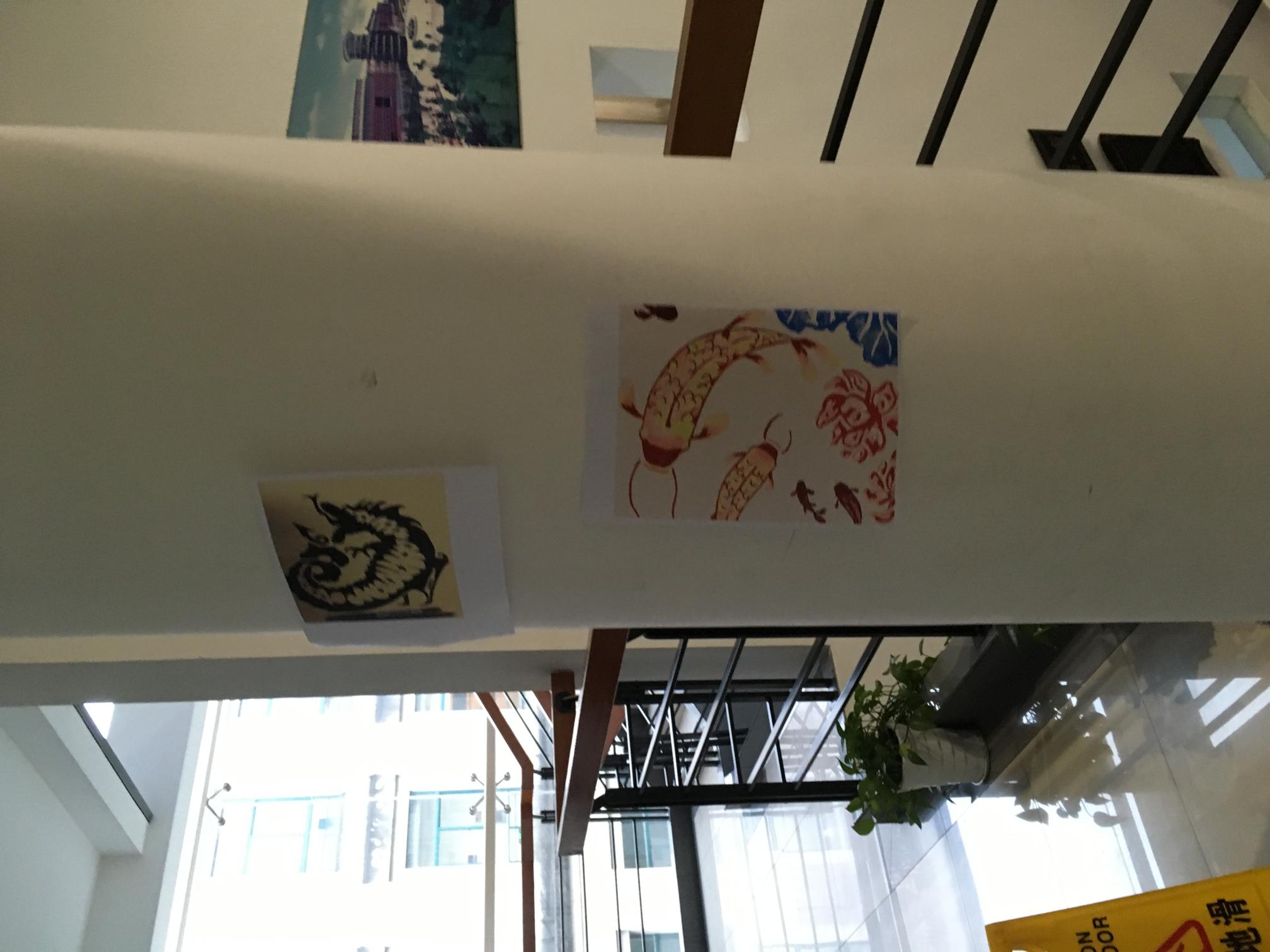}
    \end{subfigure} 
    \begin{subfigure}[t]{0.1245\textwidth}
        \includegraphics[width=\textwidth]{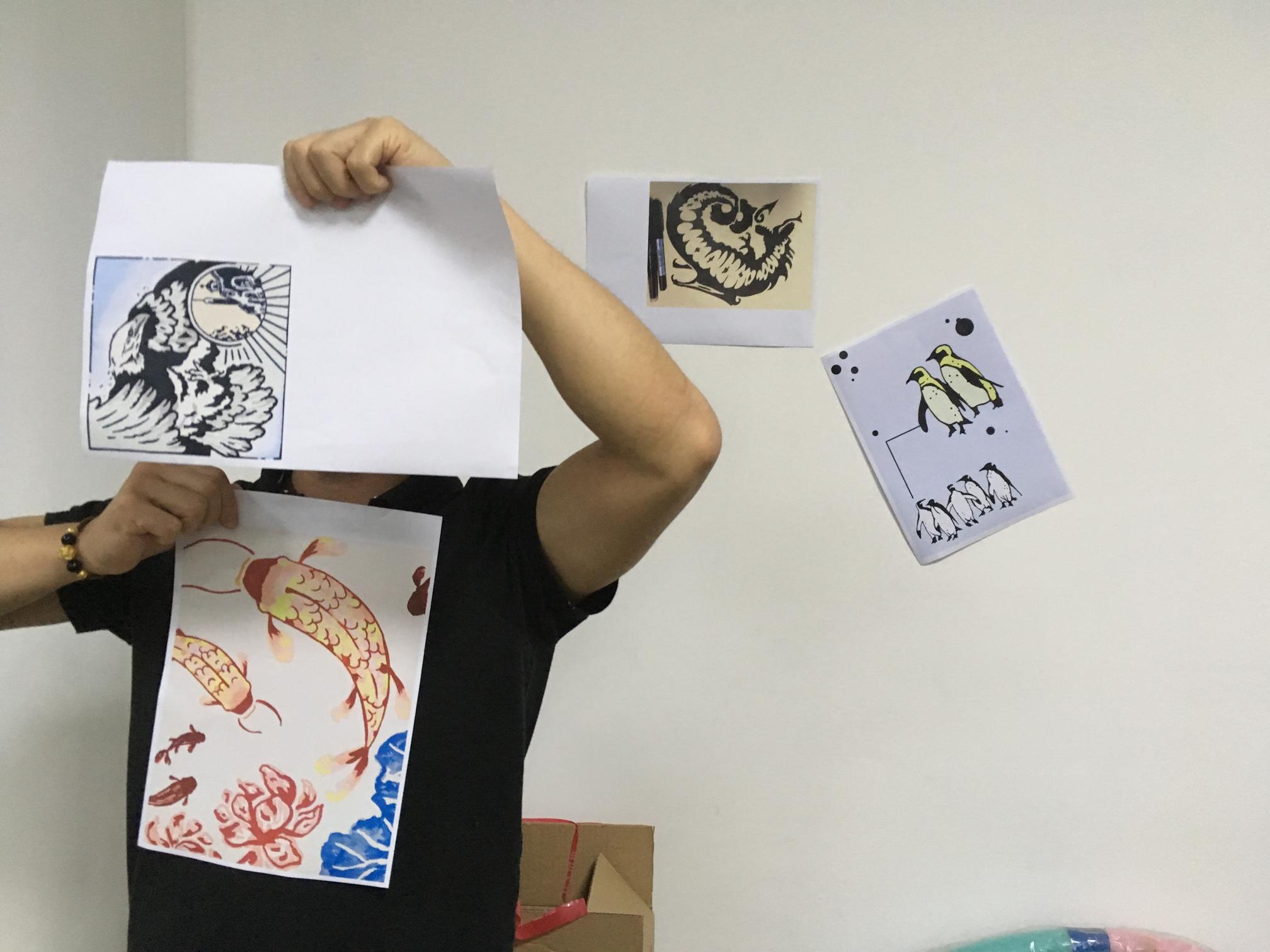}
    \end{subfigure}
    \begin{subfigure}[t]{0.095\textwidth}
        \includegraphics[width=\textwidth]{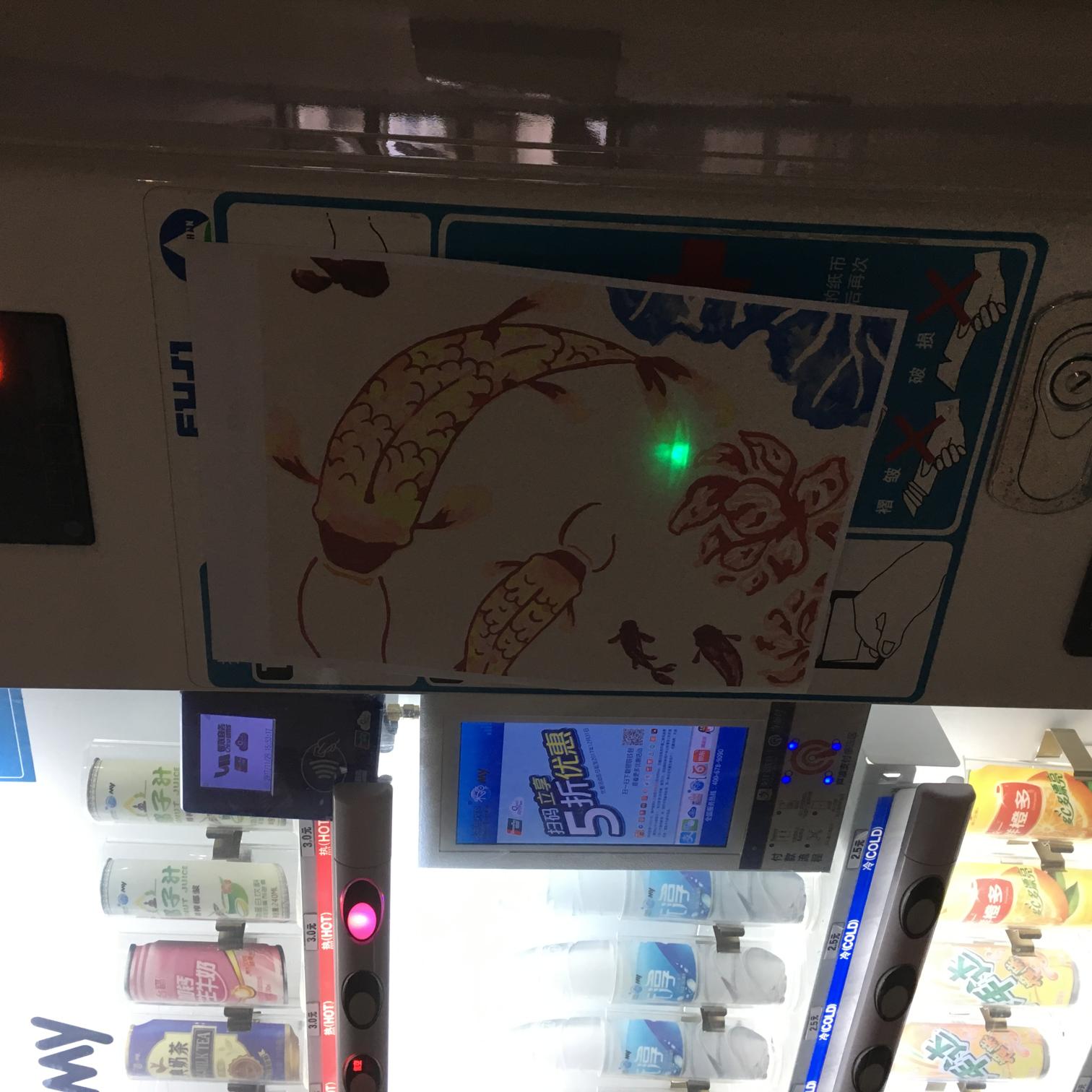}
    \end{subfigure}        
    \begin{subfigure}[t]{0.095\textwidth}
        \includegraphics[width=\textwidth]{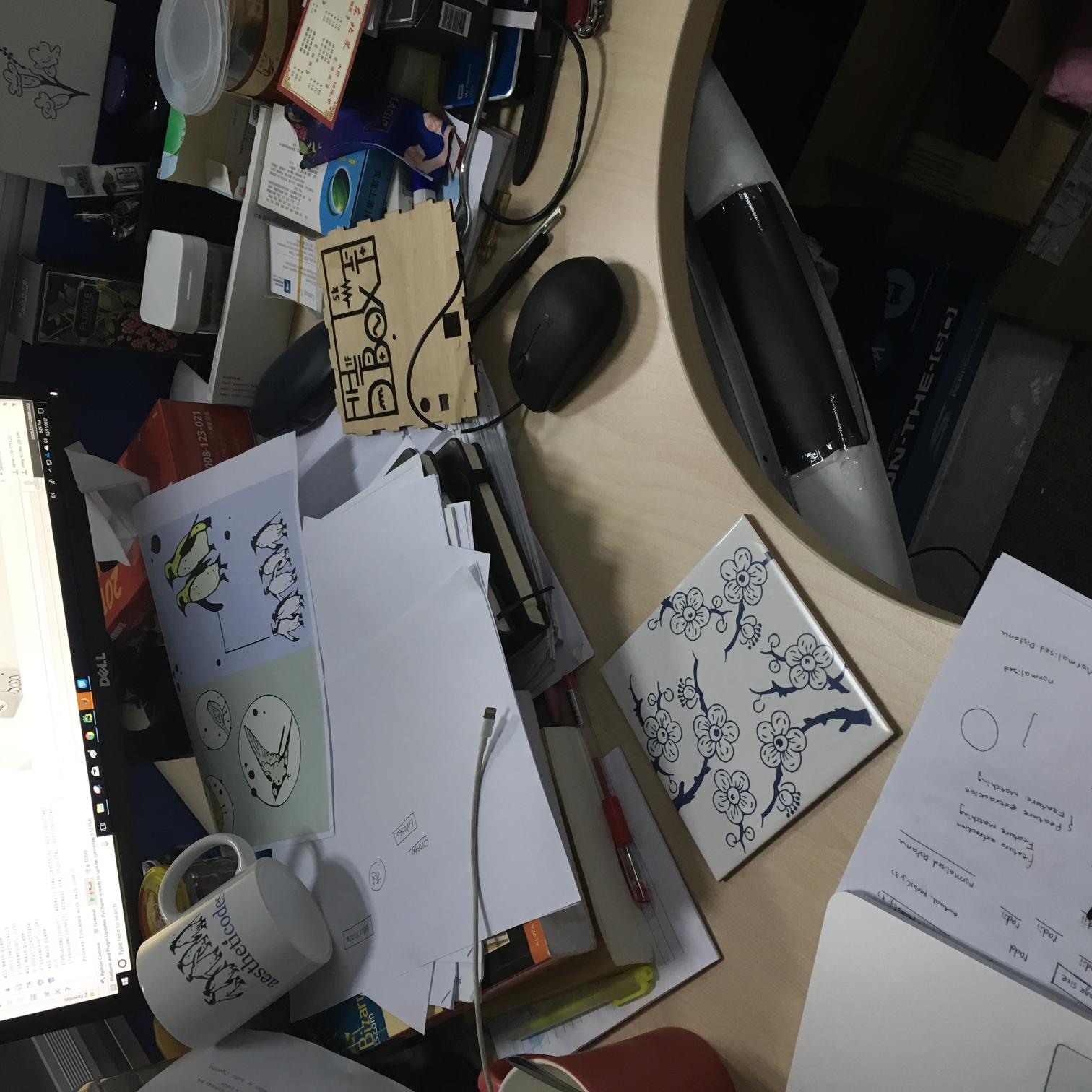}
    \end{subfigure} 
    \begin{subfigure}[t]{0.0965\textwidth}
        \includegraphics[width=\textwidth]{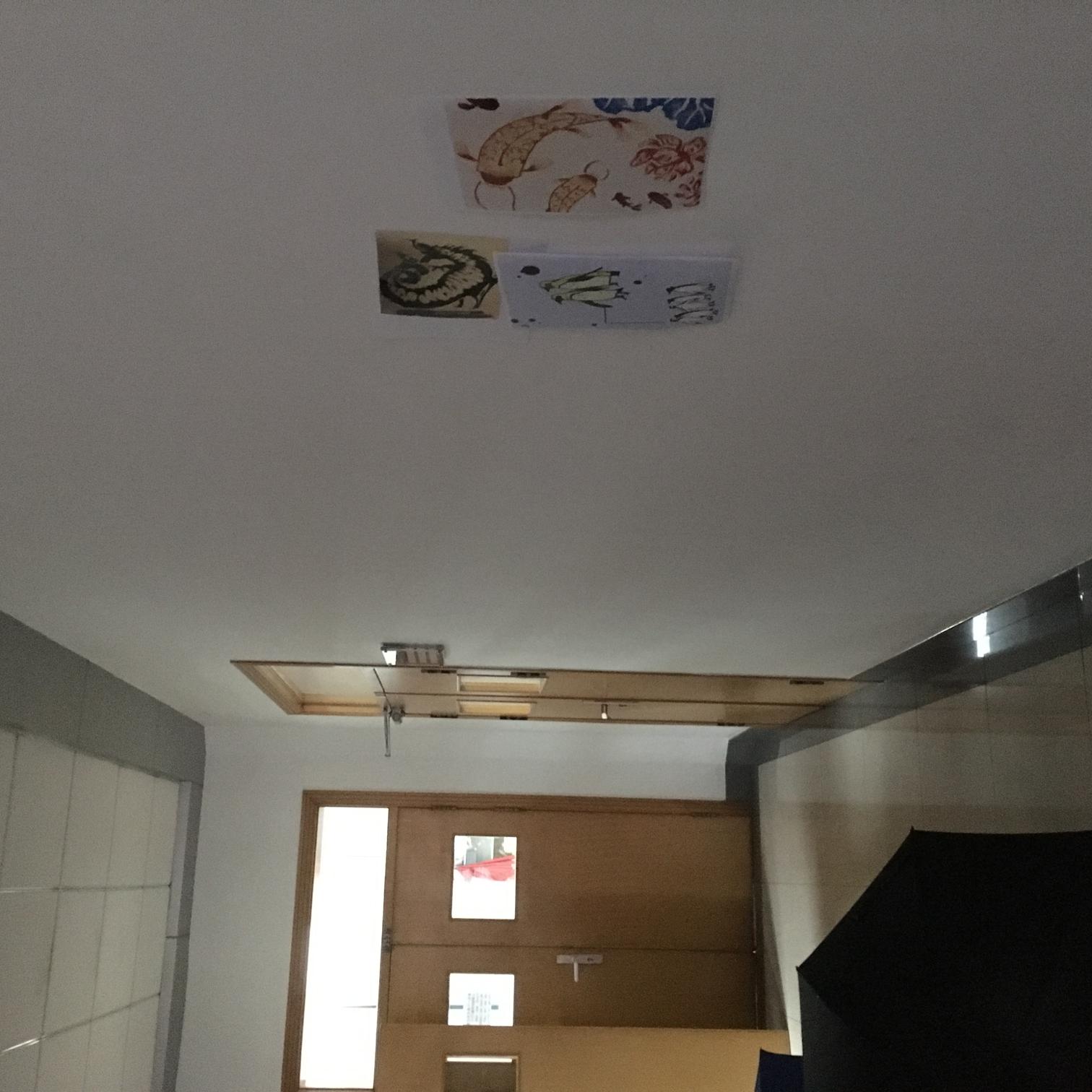}
    \end{subfigure}
    \caption{
        Example images with different ambient contexts from the SWAD dataset.
    }
    \label{fig:swad}
\end{figure*}
This section presents the experimental evaluation of the proposed {\sc VisionGuide} system.
Although heatmaps can be generated from the system, they are intermediary outputs. 
The experiments, therefore, focus on evaluating the performance of {\sc VisionGuide} on the final fine localisation---finding the centres of the Artcodes.
We first describe the dataset, performance metrics, and experimental setting used for evaluation, and then present the result in detail.


\begin{figure}[t]
    \centering  
    \begin{subfigure}[b]{0.2055\textwidth}
        \includegraphics[width=\textwidth]{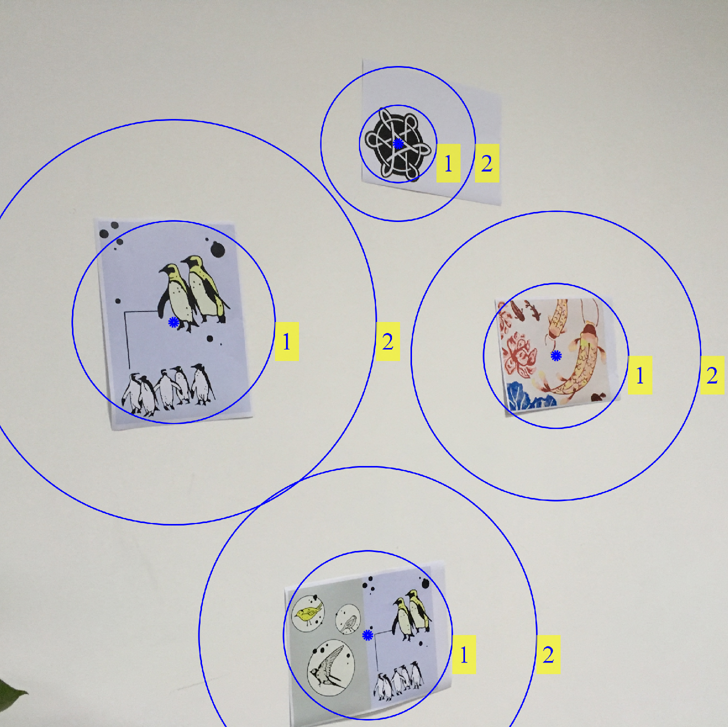}
    \end{subfigure} 
    \begin{subfigure}[b]{0.2725\textwidth}
        \includegraphics[width=\textwidth]{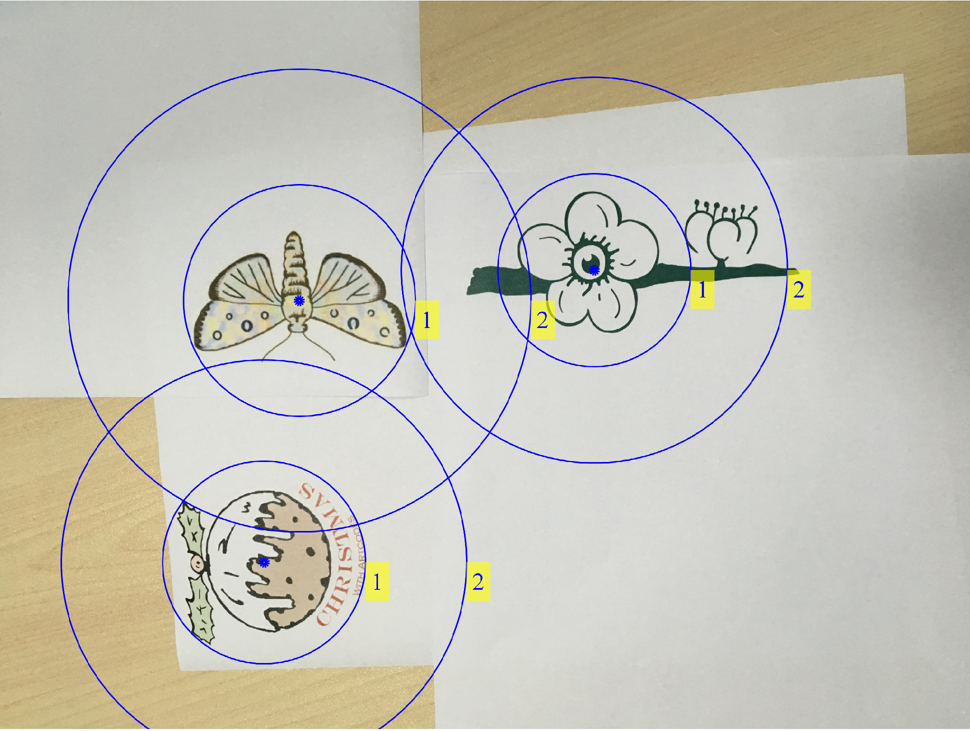}
    \end{subfigure}
    \caption{
        Illustration of normalised distance for Artcode localisation accuracy. 
        Blue circles with a radius of one normalised distance (annotated by 1) exactly cover entire Artcodes, and the ones with 2 normalised distances of two may cover Artcodes' surrounding areas. 
        The recognisable foreground of an Artcode is within one normalised distance, while its background imagery is within the range of the normalised distance of 2. 
        The ground truth location of the Artcode is the centre of the blue circles, highlighted by a blue star ({\color{blue} \textasteriskcentered}).
    }
    \label{fig:normalisedDist}
\end{figure}

\begin{figure}[t]
    \centering  
    \begin{subfigure}[b]{0.241\textwidth}
        \includegraphics[width=\textwidth]{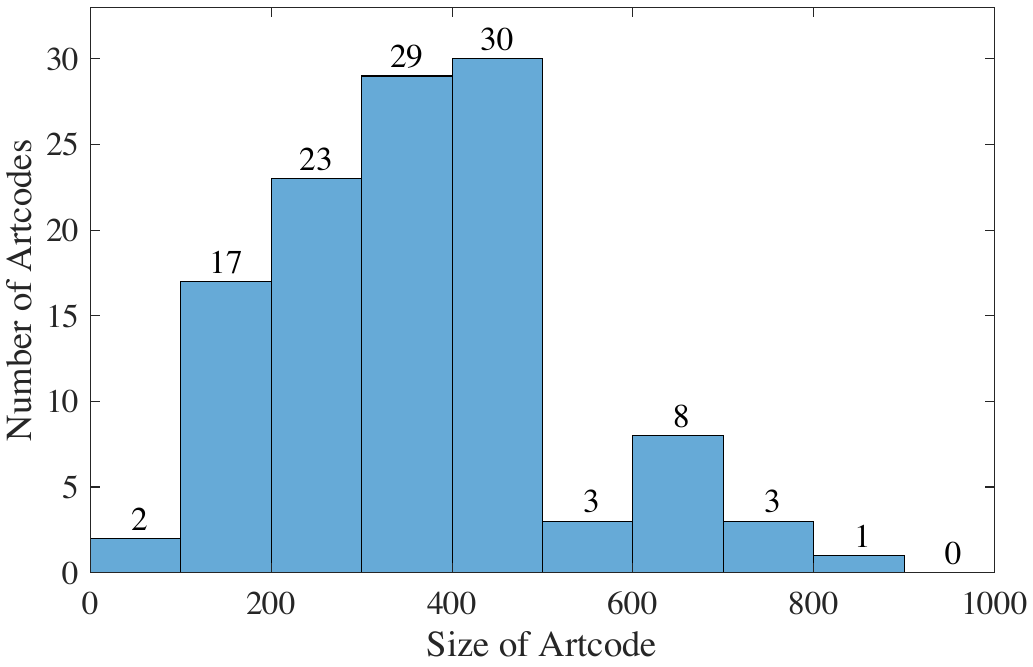}
        \caption{Bin size = 100 }
        \label{fig:ArtcodeSize_1}
    \end{subfigure} 
    \begin{subfigure}[b]{0.241\textwidth}
        \includegraphics[width=\textwidth]{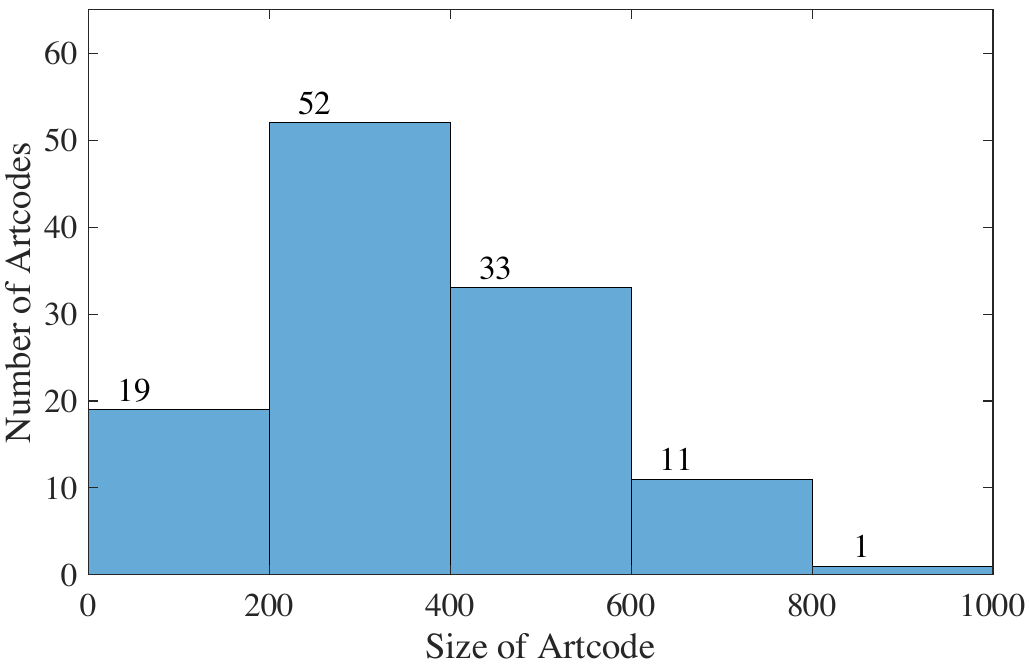}
        \caption{Bin size = 200}
        \label{fig:ArtcodeSize_2}
    \end{subfigure}
    \caption{
        Histograms of Artcode size (radii) distributions. 
        The number displayed on the top of each bin represents the total count of Artcodes falling within the size range of that particular bin.
    }
    \label{fig:ArtcodeSize}
\end{figure}

\begin{table}
\centering
\caption{Candidate parameter values for performance evaluation.}
\label{tbl:parameterValues}
    \begin{tabular}{llll} 
    \toprule
    \begin{tabular}[c]{@{}l@{}}GF~size~\\($\sigma$)\end{tabular} & \begin{tabular}[c]{@{}l@{}}HMT~threshold~\\$(h)$\end{tabular} & \begin{tabular}[c]{@{}l@{}}Sliding window\\$(w \times h)$\\\end{tabular} & \begin{tabular}[c]{@{}l@{}}Moving step\\$(step)$$(istep, jstep)$~\end{tabular}  \\ 
    \midrule
    5    & 5   & $200 \times 200$   & 20 \\
    10   & 10  & $400 \times 400$   & 40 \\
    15   & 15  & $600 \times 600$   & 60 \\
    20   & 20  & $800 \times 800$   & 80 \\
    \bottomrule
    \end{tabular}
\end{table}

\subsection{The SWAD Dataset}
\label{subsec:SWADdataset}
For experimental evaluation, we created a new dateset called the Simulated-in-the-Wild Artcodes Dataset (SWAD). 
Two main aspects were considered when creating this dataset: camera settings and ambient environment. 
Camera settings considered the camera pose and its distance to the Artcodes, while the ambient environment included lighting conditions, scene, and the positions of the Artcodes themselves.

Crafting decorated surfaces \cite{koleva2020designing, xu2017recognizing} involves considering many factors such as Artcode visual design, materials, form, function, and ambient context.
Creating a large dataset containing images of all Artcode artifacts in real-world scenarios is challenging to achieve in a reasonable timeframe.
To address these limitations, we opted for a simulated dataset containing images of Artcode visual design samples in various realistic settings. 
Artcode visual design samples were initially printed on paper and then placed in real contexts with different lighting conditions, such as offices and hallways, both in daylight and nighttime scenarios, simulating practical Artcode installations.  
A number of images containing Artcodes were captured using a consumer-grade mobile phone camera, under a wide range of camera poses.

Next, ground truth annotations of the Artcode locations were created. 
Because Artcodes have flexible geometric shapes, we used circles to annotate their locations in the images. 
The ground truth circles mainly cover the recognisable structure of the Artcode; the background imagery of the Artcode was not under consideration. 
The centre and radii of each circle denote the centroid and scale of the Artcode, respectively.  
These circles were annotated on the images at their native resolution of $4032\times3024$ or $3024\times3024$. 
Each image was accompanied with a JSON file containing the centres and radii of these circles. 
In total, we collected 44 images containing 117 Artcodes of various sizes (see example images in \autoref{fig:swad} and Artcode size distribution in \autoref{fig:ArtcodeSize}).
This dataset is publically available \footnote{\url{http://doi.org/10.5281/zenodo.2248905}}.

\subsection{Experimental Setting}\label{subsec:setting}
In this experiment, we implemented the proposed approach using MATLAB. 
For the classifier, we used the one described in \cite{xu2019artcode}, which is a random forest classifier trained on the SaTAD-4 dataset \cite{xu2017recognizing, xu2019artcode}. 
This trained classifier was employed to detect whether or not an image or sub-image contained Artcodes.

As described in \autoref{algo:calcHeatmap}, the system includes a set of parameters that need to be tuned, including the Gaussian filter size ($\sigma$), the HMT threshold ($h$), the sliding window size ($slidewin$), and the steps ($istep$ and $jstep$) for moving the sliding windows. 
To investigate the impact of each parameter individually and their combinations on the {\sc VidionGuide}, we configured the system based on the parameter values presented in \autoref{tbl:parameterValues}. 
For performance evaluation, we selected a set of metrics including precision, recall, F1 measure, and F2 measure.
Precision is the proportion of true positives out of all detections, while recall indicates the proportion of true positive detections out of all positives \cite{powers2011evaluation}. 
These two measures are often combined into F$_\beta$ measures (as defined in \autoref{eq:f_measure}), which are generally more informative than each of them individually \cite{powers2011evaluation}.
\begin{equation}\label{eq:f_measure}
    F_{\beta} = \frac{(1 + \beta^2) \cdot recall \cdot precision}{\beta^2 \cdot precision + recall} 
 \end{equation}
F1 measure ($\beta=1$), as a general measure for localisation accuracy, was adopted for performance evaluation.
As described in \autoref{subsec:artcode}, the subsequent operation after Artcode detection is decoding, indicating that detecting Artcode instances comprehensively, i.e., achieving higher recall, is more important than detection precision alone.
Therefore, we also used the F2 measure ($\beta = 2$) for evaluation, which places greater emphasis on recall than precision by giving a higher weight to {\it false negatives}.

Moreover, a distance threshold \cite{pound2017deep} was applied to select true detection. 
This threshold value can be adjusted to explore the effectiveness of the approach at varied tolerance levels. 
However, instead of directly using this distance threshold, we first normalised it based on the size of the Artcodes in each image (see \autoref{fig:normalisedDist} for an illustration). 
True positives are predictions that fall within one or two units of normalised distance from the ground truth centre, while false positives are predictions outside this range.
As presented in \autoref{fig:normalisedDist}, the areas specified by the normalised distance of different Artcodes are allowed to overlap. 
Predicted points within the overlapped area are assigned as the predicted location of the Artcode with the smaller normalised distance to this point. 
Thus, a predicted point that is geometrically further from the Artcode than another predicted point is likely to be assigned to this Artcode as its predicted location due to its smaller normalised distance to this point.

\subsection{Results}
\label{subsec:results}
\begin{figure*}[t]
    \centering  
    \begin{subfigure}[b]{0.245\textwidth}
        \includegraphics[width=\textwidth]{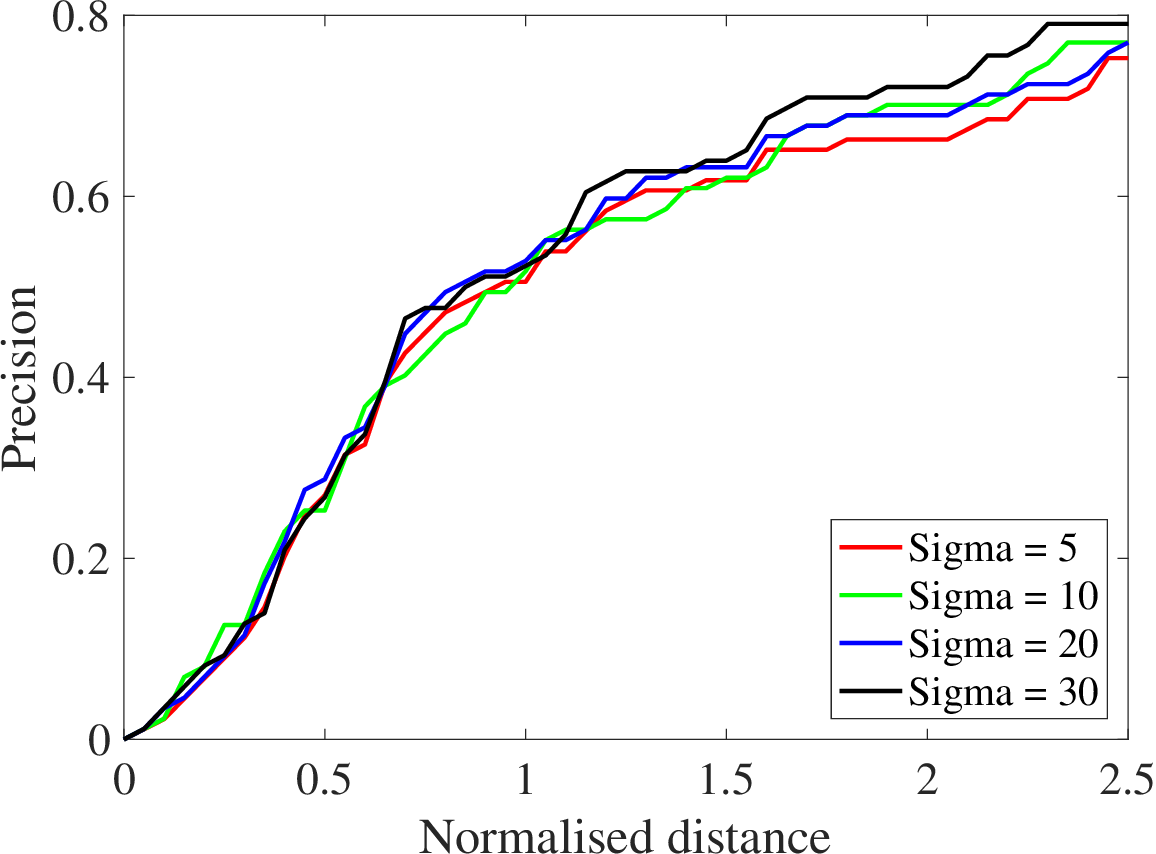}
        \caption{Precision}
        \label{fig:precision-GF}
    \end{subfigure} 
    \begin{subfigure}[b]{0.245\textwidth}
        \includegraphics[width=\textwidth]{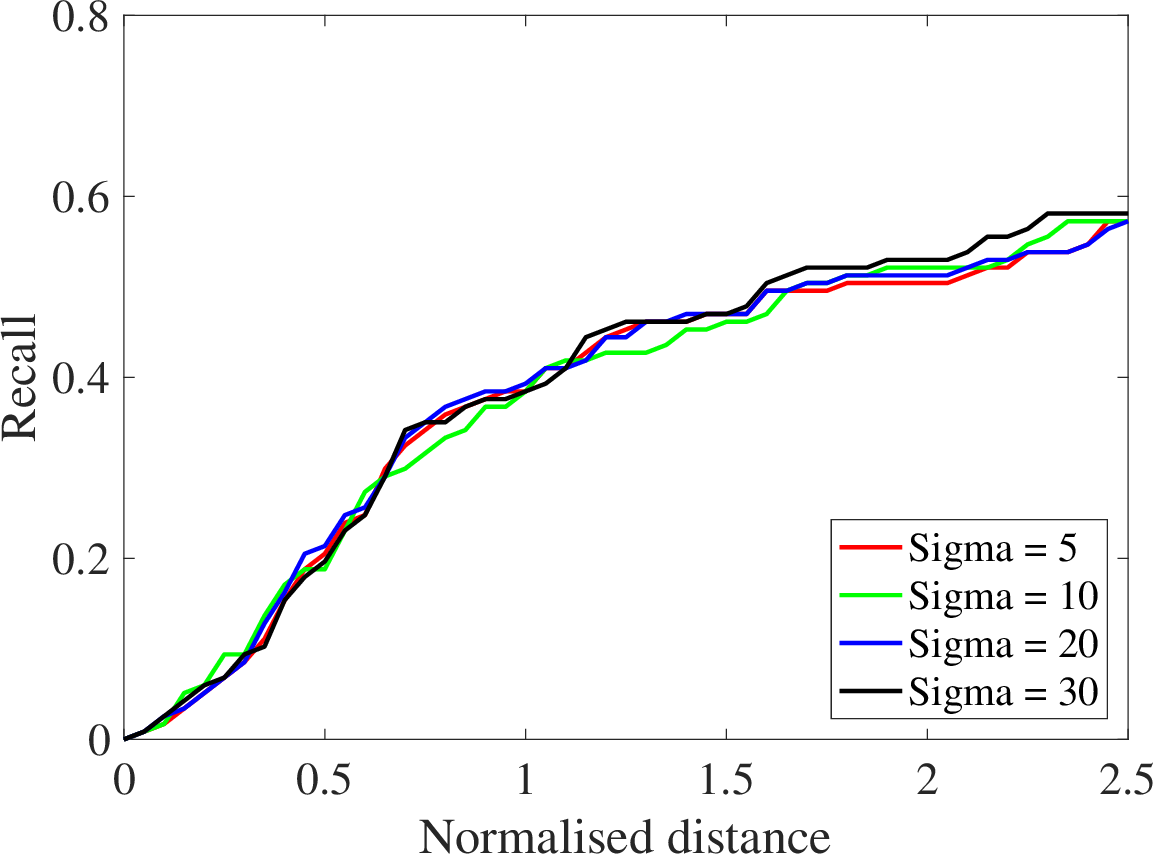}
        \caption{Recall}
        \label{fig:recall-GF}
    \end{subfigure}
     \begin{subfigure}[b]{0.245\textwidth}
        \includegraphics[width=\textwidth]{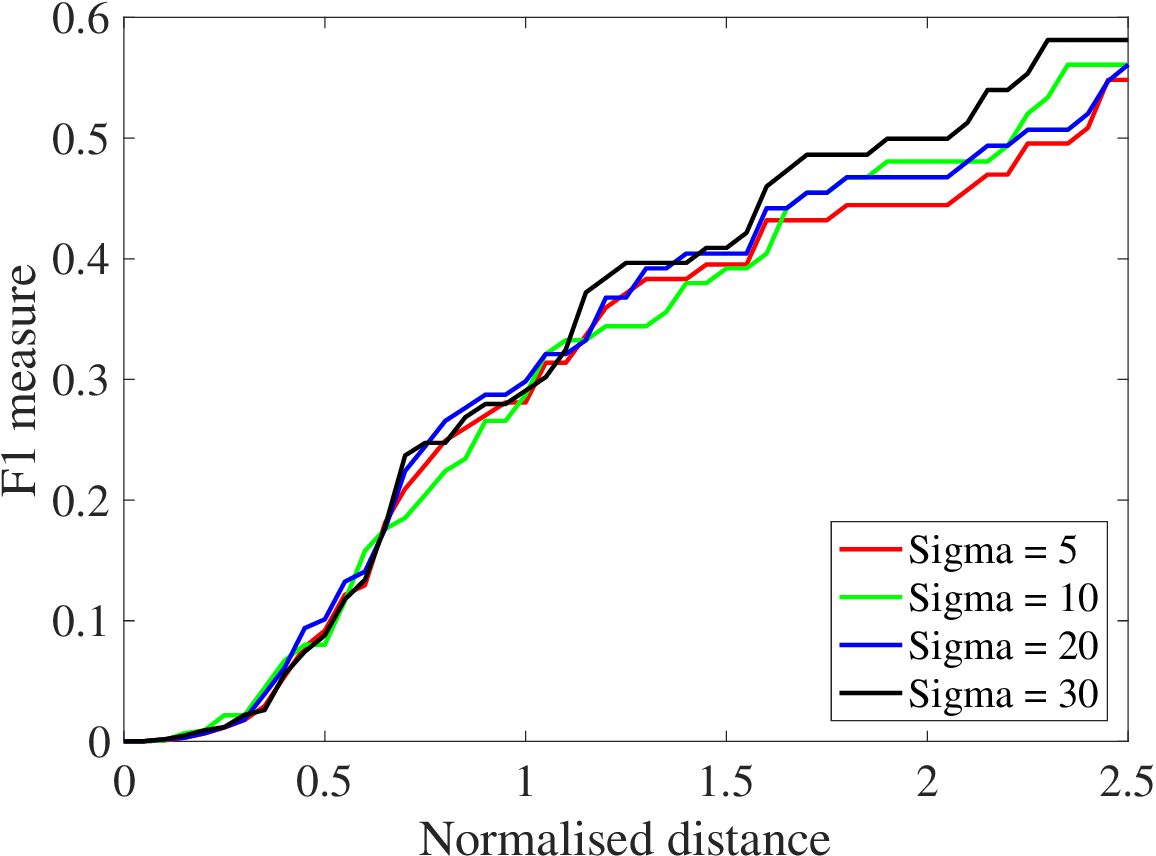}
        \caption{F1 measure}
        \label{fig:fbeta-GF}
    \end{subfigure}
    \begin{subfigure}[b]{0.245\textwidth}
        \includegraphics[width=\textwidth]{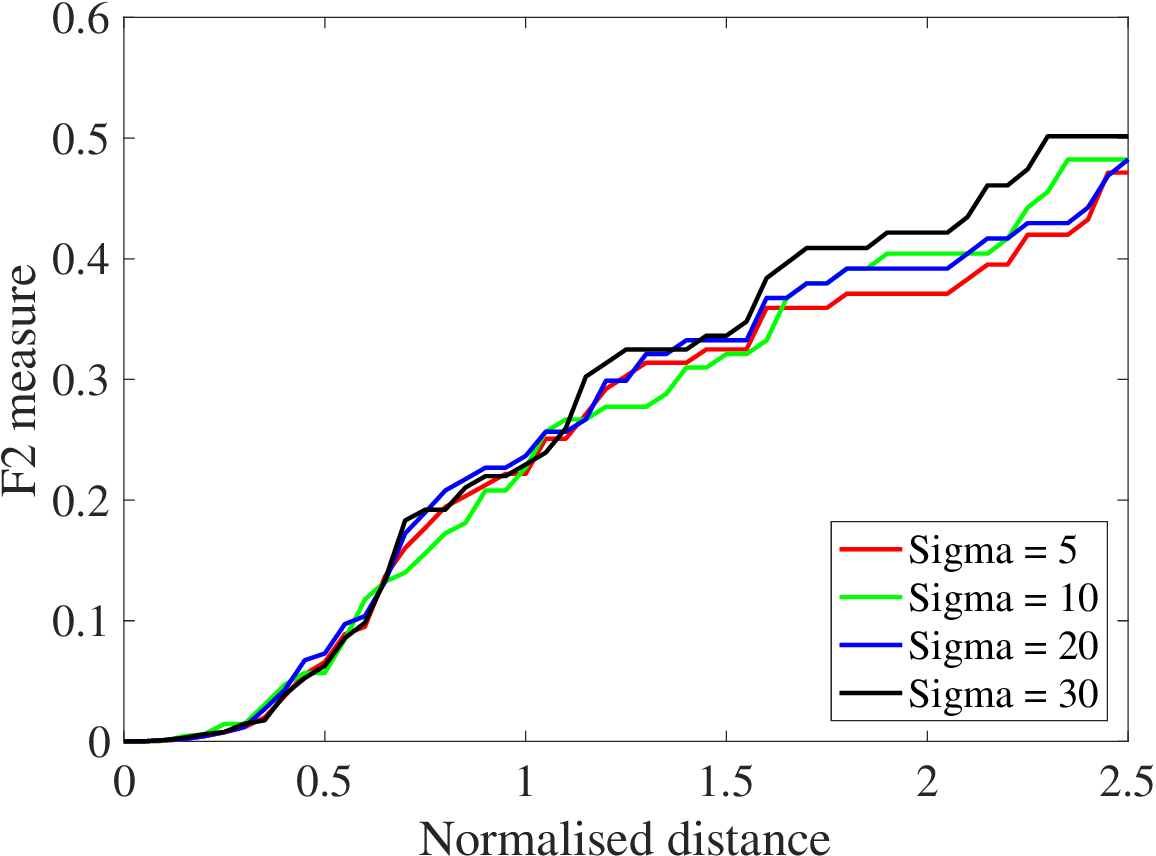}
        \caption{F2 measure}
        \label{fig:fbeta-GF}
    \end{subfigure}
    \caption{
        Performance comparison of {\sc VisionGuide} with Gaussian filters of different sizes (other parameters: $slidewin = 800 \times 800$, $step = 40$, $h = 10$).}
    \label{fig:result-GF}
\end{figure*}

\begin{figure*}
    \centering  
    \begin{subfigure}[b]{0.245\textwidth}
        \includegraphics[width=\textwidth]{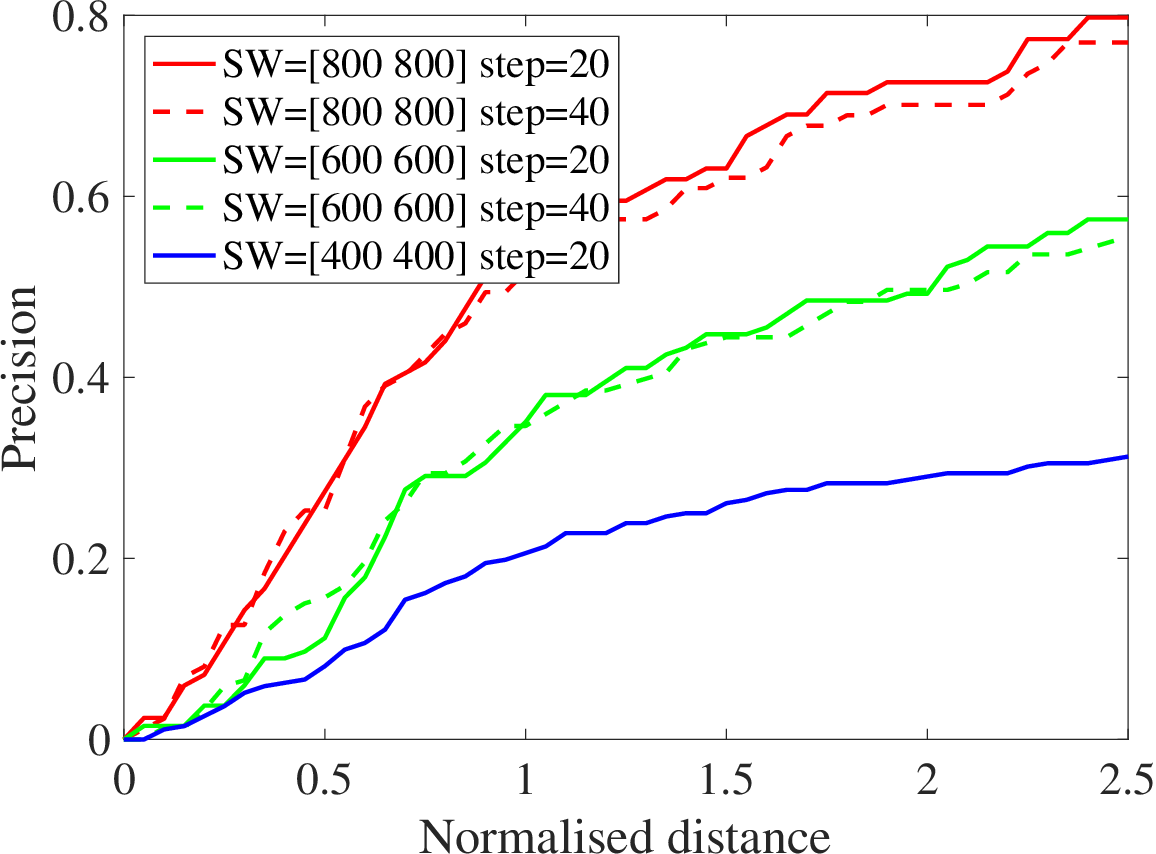}
        \caption{Precision}
        \label{fig:precision-SW}
    \end{subfigure} 
    \begin{subfigure}[b]{0.245\textwidth}
        \includegraphics[width=\textwidth]{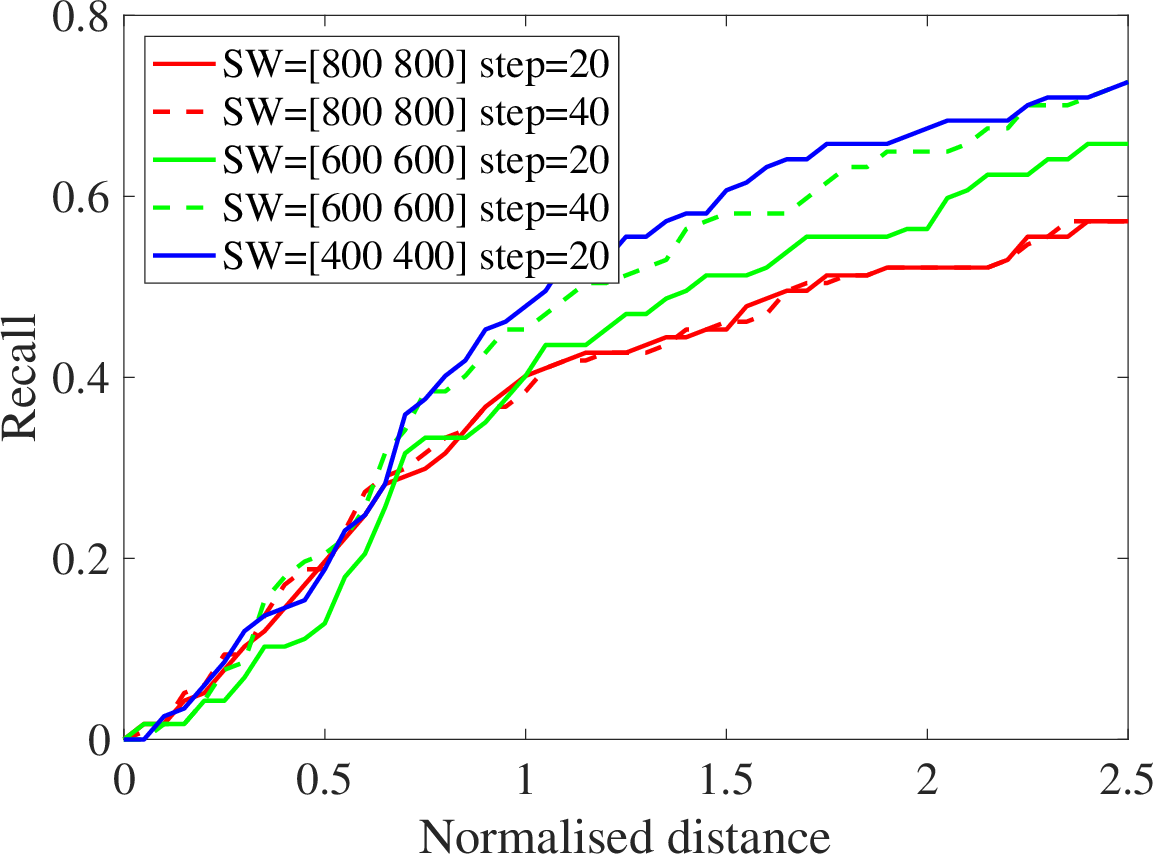}
        \caption{Recall}
        \label{fig:recall-SW}
    \end{subfigure}
    \begin{subfigure}[b]{0.245\textwidth}
        \includegraphics[width=\textwidth]{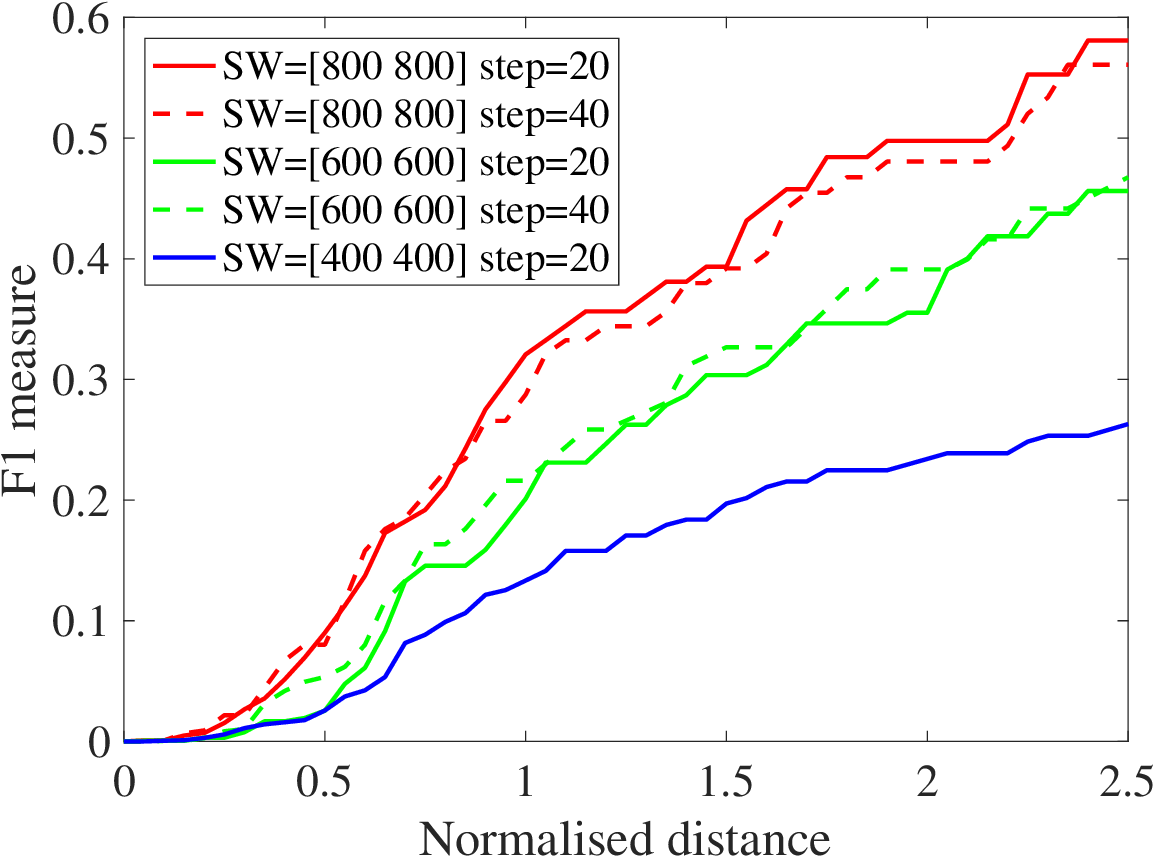}
        \caption{F1 measure}
        \label{fig:f1-SW}
    \end{subfigure}
    \begin{subfigure}[b]{0.245\textwidth}
        \includegraphics[width=\textwidth]{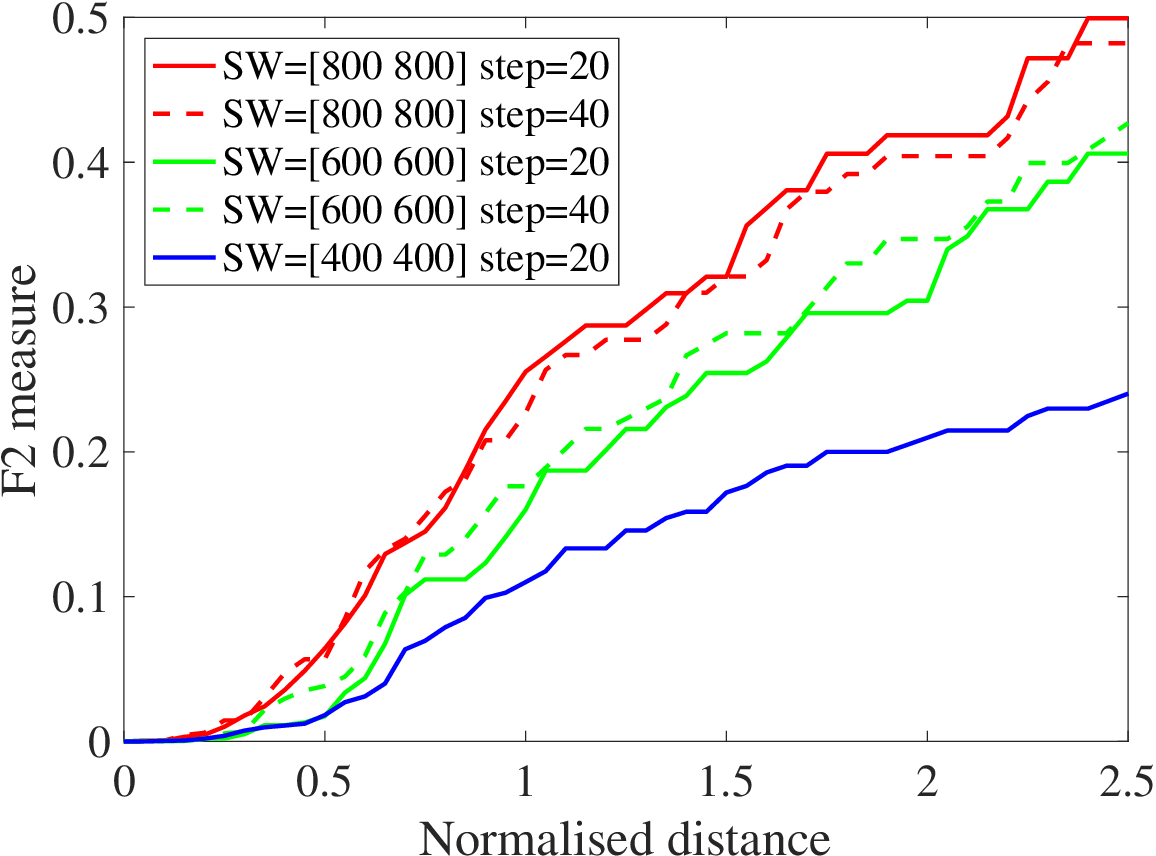}
        \caption{F2 measure}
        \label{fig:f2-SW}
    \end{subfigure}
    \caption{
        Performance comparison of {\sc VisionGuide} with different sizes of slide windows and lengths of the moving step (other parameters: $\sigma = 10$, $h = 10$).
    }
    \label{fig:result-SW}
\end{figure*}

\begin{figure*}
    \centering  
    \begin{subfigure}[b]{0.245\textwidth}
        \includegraphics[width=\textwidth]{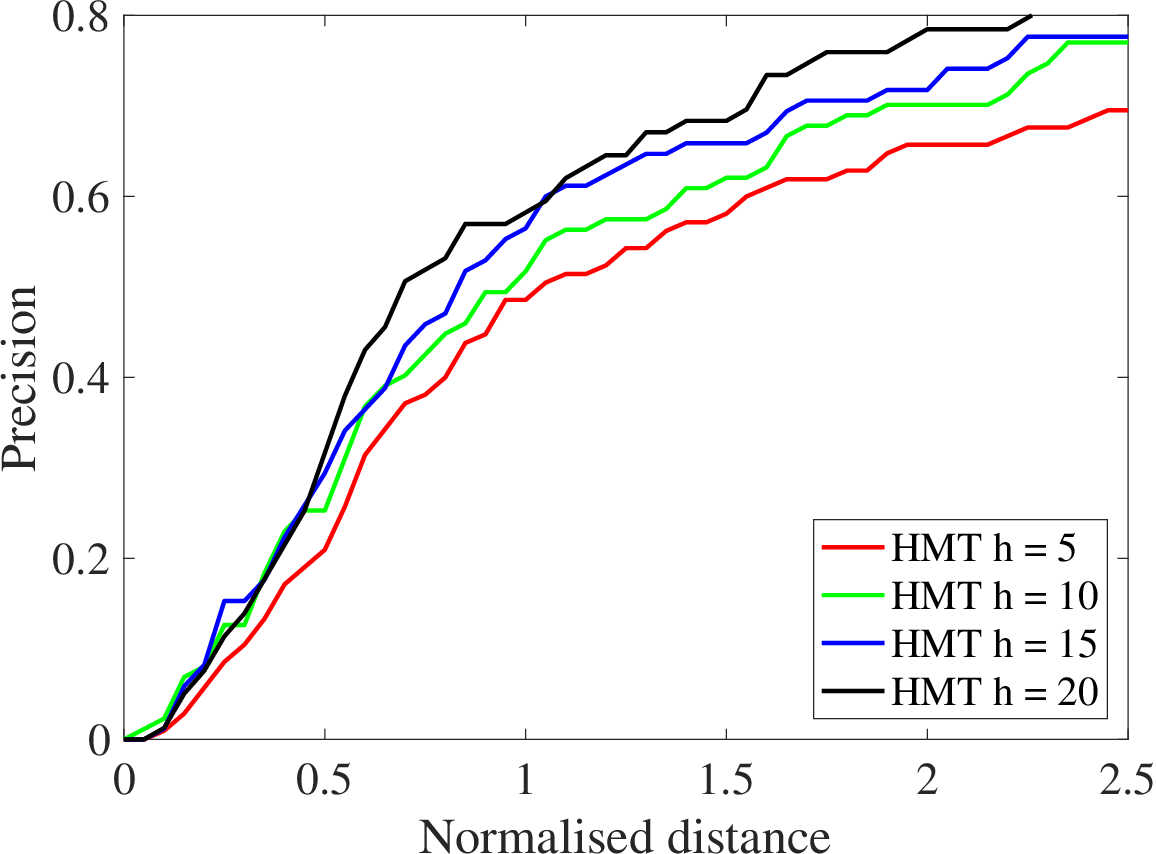}
        \caption{Precision}\label{fig:precision-HMT}
    \end{subfigure} 
    \begin{subfigure}[b]{0.245\textwidth}
        \includegraphics[width=\textwidth]{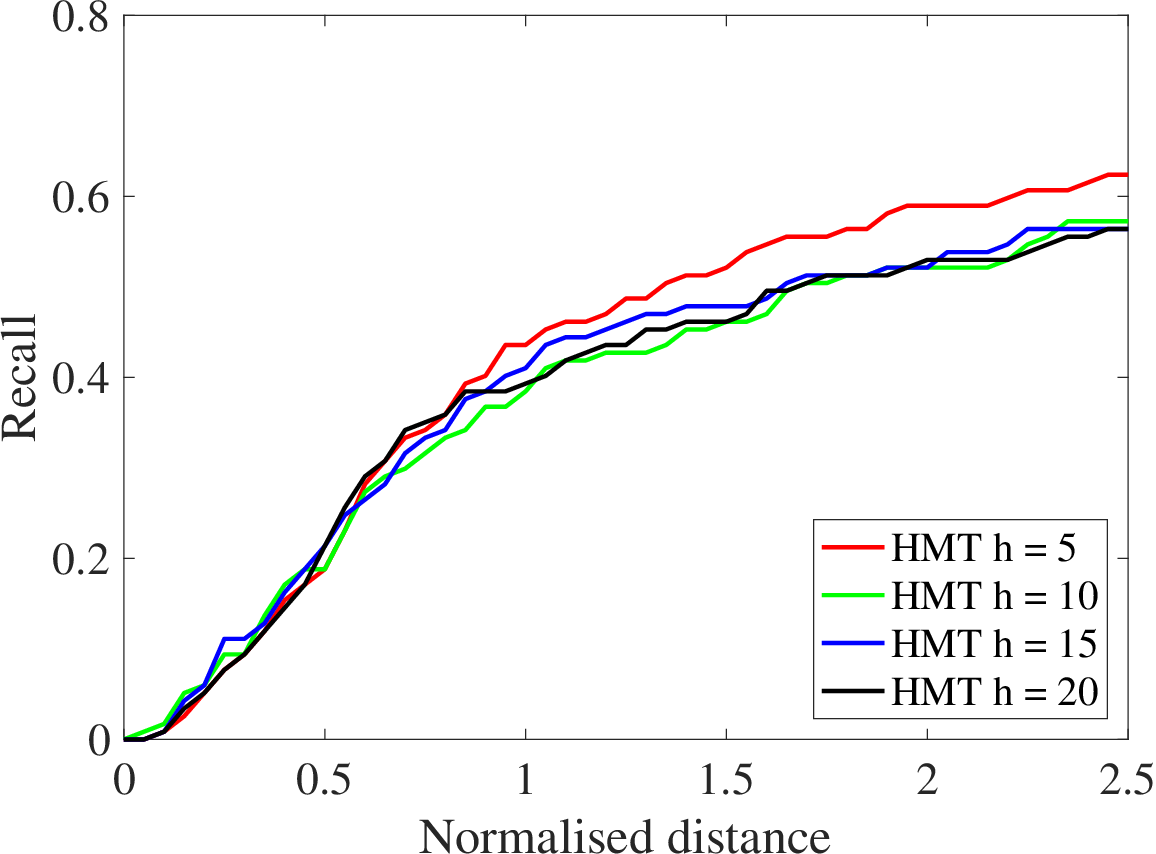}
        \caption{Recall}\label{fig:recall-HMT}
    \end{subfigure}
     \begin{subfigure}[b]{0.245\textwidth}
        \includegraphics[width=\textwidth]{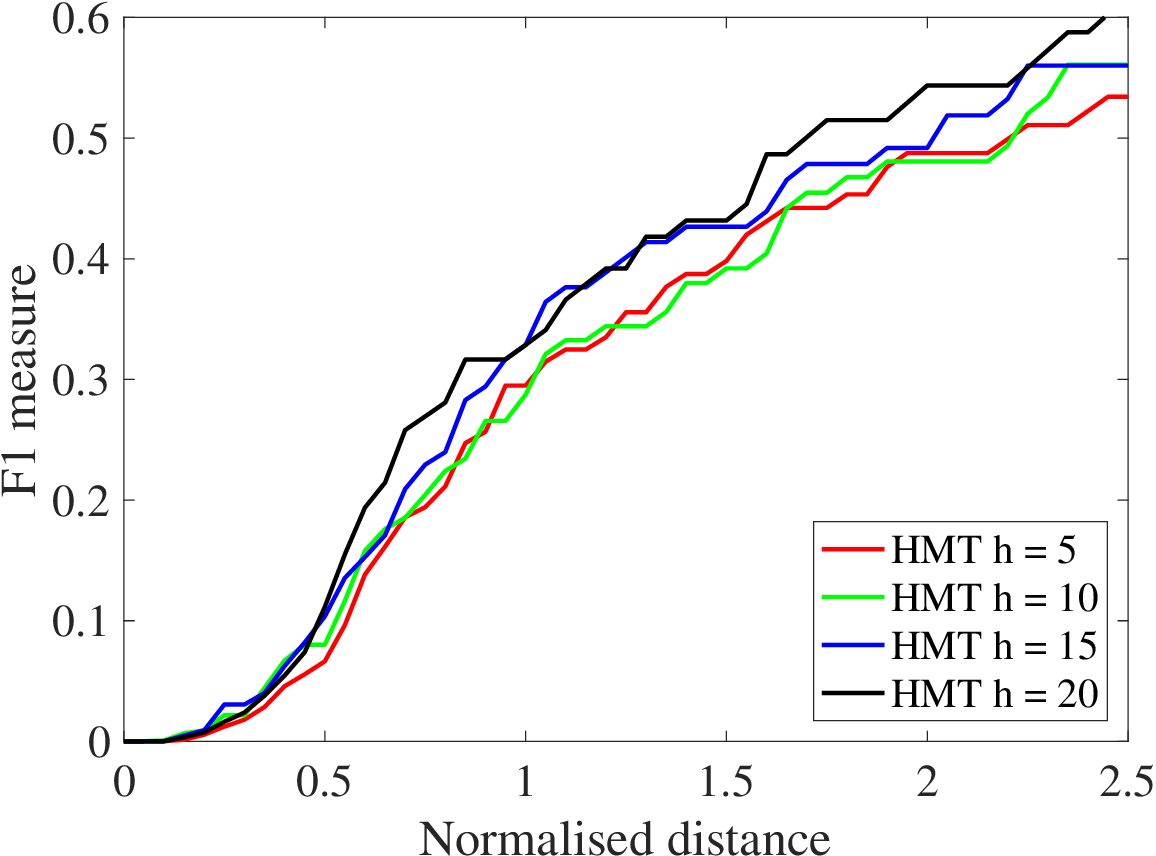}
        \caption{F1 measure}\label{fig:f1-HMT}
    \end{subfigure}
    \begin{subfigure}[b]{0.245\textwidth}
        \includegraphics[width=\textwidth]{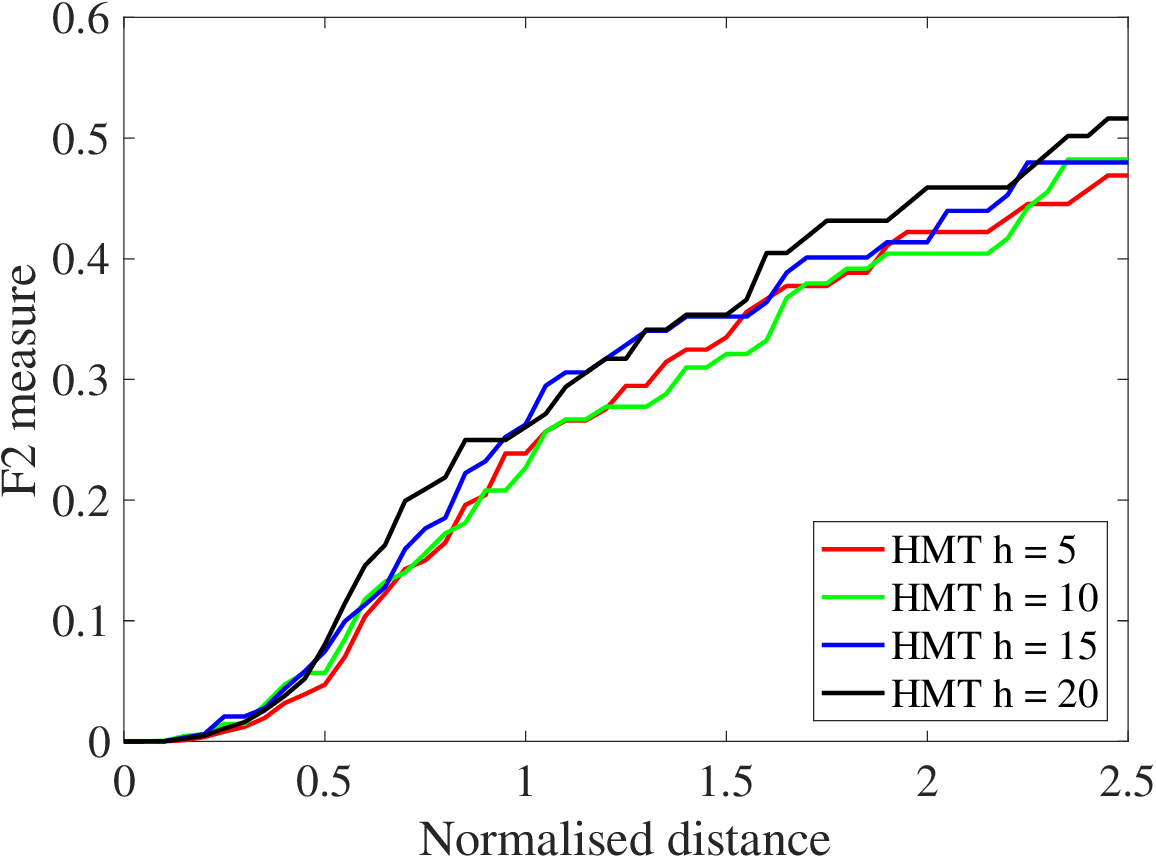}
        \caption{F2 measure}\label{fig:f2-HMT}
    \end{subfigure}
    \caption{
        Performance comparison of {\sc VisionGuide} with different values of threshold level of H-Maxima transformation (other parameters: $slidewin=800 \times 800$, $step=40$, $\sigma=10$).
    }
    \label{fig:result-HMT}
\end{figure*}

The performance of {\sc VisionGuide} was evaluated on the SWAD dataset using the four performance metrics.
True positives and false negatives were determined based on whether or not the predicted locations were within a given threshold of normalised distance (1 or 2) from the ground truth locations.
Specifically, if the calculated normalised distance between the predicted and ground truth location was less than or equal to the given threshold, then the prediction was considered correct. 
Therefore, varying this threshold of normalised distance can affect the evaluation metrics: a lower threshold results in lower value of the evaluation metrics. 
Experiments were conducted to study the impact of the tuning parameters described in 
 \autoref{tbl:parameterValues}, leading to the results shown in \autoref{fig:result-GF}, \autoref{fig:result-SW}, and \autoref{fig:result-HMT}.

\subsubsection{Size of Gaussian Filter}
Experiment was conducts to study the impact of the Gaussian filter size ($\sigma$), with results shown in \autoref{fig:result-GF}. 
$\sigma$ determines the kernel size of a Gaussian filter, representing its width and height.
The filter size is calculated as follows:
\begin{equation}\label{eq:gaussSize}
    size = 2 \times \ceil*{2 \times \sigma} + 1
\end{equation}
where the function $\ceil{x}$ rounds the value of $x$ up to the nearest integer greater than or equal to $x$. 
For example, if $\sigma=10$, the size of the Gaussian filter would be $2\times\ceil{2\times10}+1=2\times20+1=41$.

The impact of Gaussian filters of different kernel sizes on {\sc VisionGuide}'s performance was compared by varying the values of $\sigma$ while keeping the values of the other parameters unchanged. 
In this experiment, the values of $\sigma$ were selected from $[5, 10, 20, 30]$, and the other parameters were assigned fixed values: $slidewin=800\times800$; $step=40$; and $h=10$. 
The experimental results are presented in \autoref{fig:result-GF}. 
It can be seen that {\sc VisionGuide} achieves a very similar perfomance overall under the Gaussian filters of the four sizes from a normalised distance of 0 to 1.5, in terms of precision, recall, F1 and F2 measures. 
This similarity changes when the normalised distance is greater than 1.5: {\sc VisionGuide} with $\sigma = 30$ achieves higher values, indicating that more locations of Artcodes further away from the ground truth center of the Artcodes were correctly predicted. 
However, larger Gaussian filters do not consistently result in better localisation performance. 
Smaller Gaussian filters ($\sigma = 10$) obtained better evaluation values in terms of the four metrics than larger Gaussian filters (such as $\sigma = 20$), and both appeared to perform better than the smallest Gaussian filter with a kernel size of 5.

As discussed in \autoref{par:gaussFiltering}, the selection of the Gaussian filter size is related to the size of the moving step. 
This relationship is confirmed by the results of the next experiment (see \autoref{fig:result-SW}). If the length of the moving step in this experiment was set to 40, according to \autoref{eq:gaussSize}, the smallest $\sigma$ should be 10. 
Larger Gaussian filters (such as $\sigma = 20$) can fill and smooth the valleys, but they may not necessarily perform better; in fact, they may even perform worse. 
However, much larger Gaussian filters ($\sigma = 30$), whose kernel size reaches 121, obtain better performance overall. 
The improved performance is attributed to the correct predictions that are far away from the centres of the Artcodes. 
It is important to note that larger Gaussian filters incur higher computational costs. 
Hence, approximately one-quarter of the length of the moving step should be an appropriate choice for Gaussian filter size $\sigma$.

\subsubsection{Size of Sliding Window and Moving Step}
The next experiment investigated the impact of the size of the sliding window ($slidewin$) and its moving step ($istep$ and $jstep$). 
Because the size of the sliding window and its moving step are closely related, both determine the accumulated score of the output heatmap at point $(x, y)$ (see \autoref{eq:accumulatedScore}). 
We compared the performance of {\sc VisionGuide} with different combinations of $slidewin$ and $step$, with the results presented in \autoref{fig:result-SW}. 
{\sc VisionGuide} with $slidewin = 800 \times 800$ and $step = 20$ achieves better performance than with $step = 40$, and other combinations of $slidewin$ and $step$, across all the normalised distances. 
The length of the moving step directly determines the total number of sub-images that need to be classified. 
The larger the moving step, the smaller the number of sub-images, and the lower the computational cost.  
Hence, when the moving step is 20, twice the number of sub-images need to be classified compared to the moving step 40. 
As shown in \autoref{fig:precision-SW}, \autoref{fig:f1-SW} and \autoref{fig:f2-SW}, {\sc VisionGuide} with $slidewin = 800 \times 800$ and $step = 20$ is slightly better than that of $slidewin = 800 \times 800$ and $step = 40$. 
However, {\sc VisionGuide} with $step = 20$ incurs double the computational cost compared with $step = 40$. 
Considering the computational cost, the combination of $slidewin = 800 \times 800$ and $step = 40$ could be a more desirable choice for the size of the sliding window and the moving step.

Overall, {\sc VisionGuide} with a larger sliding window shows better performance in terms of precision, F1 measure, and F2 measure. 
The selection of the size of the sliding window is relevant to the size of the Artcodes in the dataset. 
A larger Artcode requires larger sliding windows. 
However, an image could contain Artcodes of different sizes. 
In this experiment, only a fixed size sliding window was used, and a sliding window that is larger than most Artcodes would be an appropriate choice. 
\autoref{fig:ArtcodeSize} shows the distributions of Artcode sizes in the SWAD dataset. 
It can be seen that the sizes of most Artcodes are up to $800 \times 800$. 
A $400 \times 400$ window could only cover approximately half of the Artcodes. 
This distribution partially explains the better performance of {\sc VisionGuide} with a $800 \times 800$ sliding window.

In terms of recall, {\sc VisionGuide} with a smaller sliding window exhibits better performance compared to using a larger sliding window.
Specifically, {\sc VisionGuide} with the smallest sliding window ($slidewin = 400 \times 400$) achieves the highest recall value. 
This higher recall but lower precision suggests that {\sc VisionGuide} with $slidewin = 400 \times 400$ predicts a greater number of false positives than its counterpart with a larger sliding window, indicating {\sc VisionGuide} with a small size of sliding window is sensitive to the background objects. 
The size of the moving step also plays a role in the performance. 
As shown in \autoref{fig:result-SW}, {\sc VisionGuide} with $slidewin= 600 \times 600$ and $step = 40$ outperforms its counterpart with a smaller moving step ($step = 20$) across all four metrics. 
However, this behaviour is not observed with $slidewin = 800 \times 800$, in which the larger moving step performs better than a smaller one. 
Therefore, careful selection of both the sliding window size and the length of the moving step is crucial for the performance of {\sc VisionGuide}. 
Experiments show that {\sc VisionGuide} with $slidewin = 800 \times 800$ and $step = 40$ demonstrates reasonably good performance in terms of both evaluation metrics computational cost. 
Consequently, this combination of sliding window size and moving step length was used in the subsequent experiments to study the impact of the other parameters.

\subsubsection{HMT Threshold}
An experiment was conducted to study the impact of the HMT threshold ($h$) on {\sc VisionGuide}'s performance. 
The experimental results, with different values of $h$ ($h = [5, 10, 15, 20]$), are presented in \autoref{fig:result-HMT}. 
Overall, {\sc VisionGuide} with $h = 20$ achieves the best performance in terms of precision, F1 measure, and F2 measure. 
This indicates that larger values of $h$ lead to better performance across these three metrics. 
However, {\sc VisionGuide} with the smallest value of $h$ ($h = 5$) obtains a higher recall value, while its counterparts with other values of $h$ are very close, as shown in \autoref{fig:recall-HMT}.

As discussed in \autoref{sec:hmaxTransformation}, the tuning of the HMT threshold ($h$) plays a pivotal role in determining the size and number of regional maxima during the peak finding phase: the larger the value of $h$, the larger the size of regional maxima, and the smaller the number of regional maxima. 
{\sc VisionGuide} with larger values of $h$ favours the selection of higher peaks, indicative of more sub-images covered by these peak regions being predicted as Artcodes, thereby increasing the likelihood of identifying Artcode-associated peaks.
Additionally, the increased size of regional maxima facilitates more precise calculations of Artcode positions (i.e., centroids of regional maxima), contributing to reduced normalised distances.
Employing a larger $h$ value in the H-Maxima transformation may eliminate lower peaks, as depicted in \autoref{fig:hmaxtransform-graphic}.

The results of these three experiments show that the proposed approach---{\sc VisionGuide}---achieves overall fairly good performance in fine localisation of Artcodes. 
This achievement is particularly noteworthy considering the inherent difficulty associated with detecting visual elements such as Artcode patterns due to their flexible shapes and tendency to blend into their surrounding ambient environment.

\section{Discussion and Examination}
\label{sec:discussion}
\begin{figure*}
    \centering     
    \begin{subfigure}[t]{0.245\textwidth}
        \includegraphics[width=\textwidth]{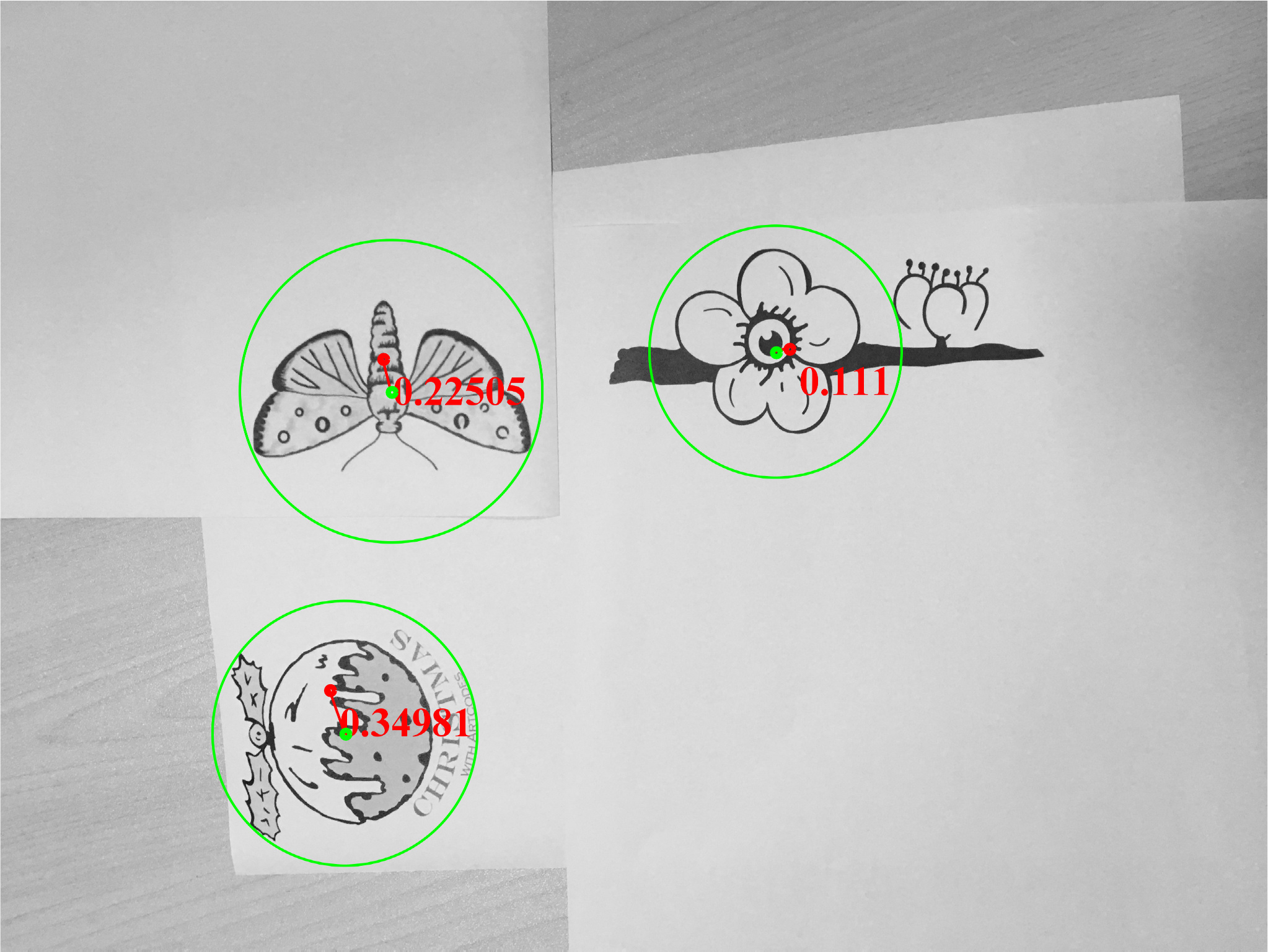}
    \end{subfigure} 
     \begin{subfigure}[t]{0.245\textwidth}
        \includegraphics[width=\textwidth]{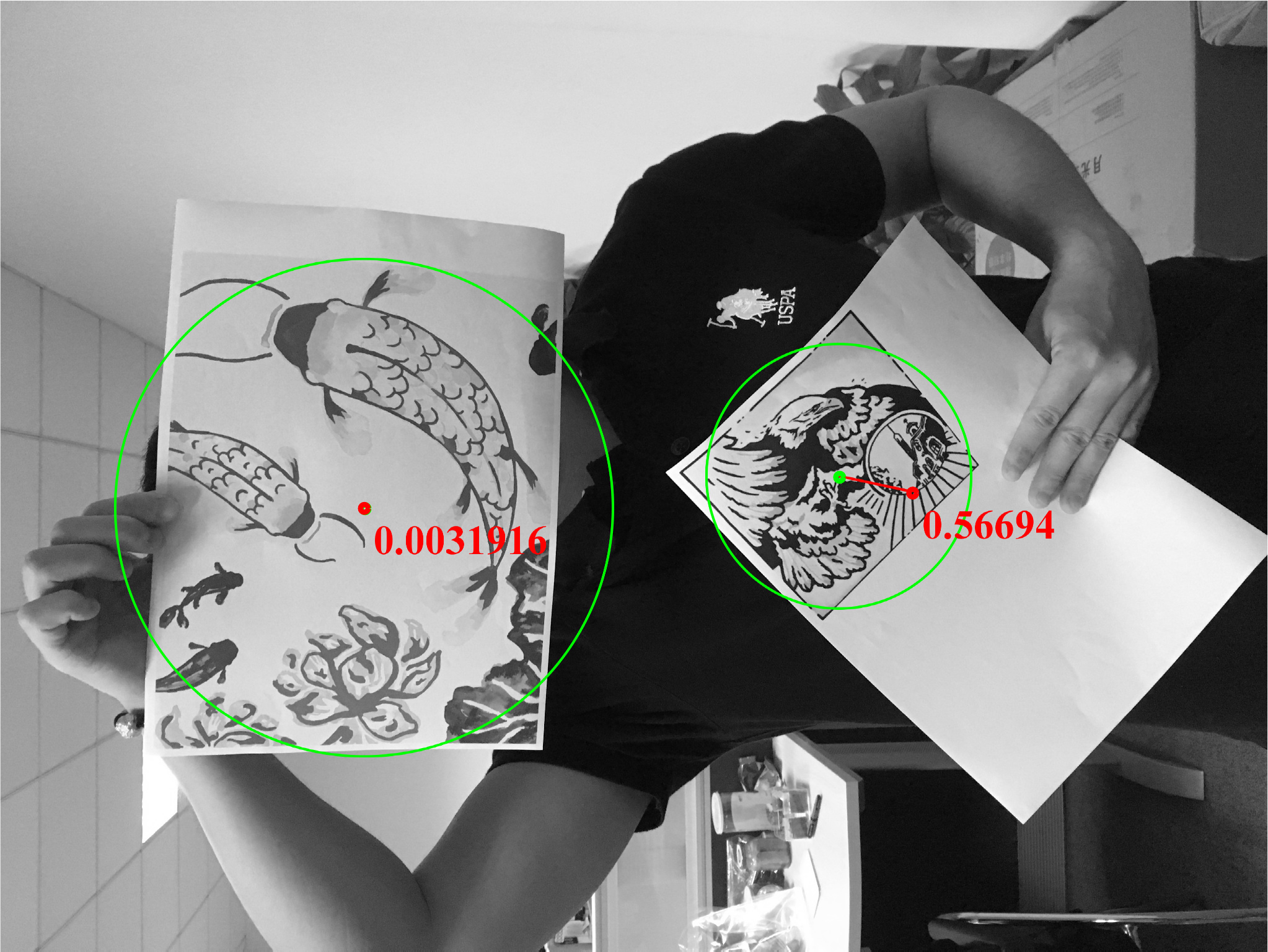}
    \end{subfigure} 
    \begin{subfigure}[t]{0.245\textwidth}
        \includegraphics[width=\textwidth]{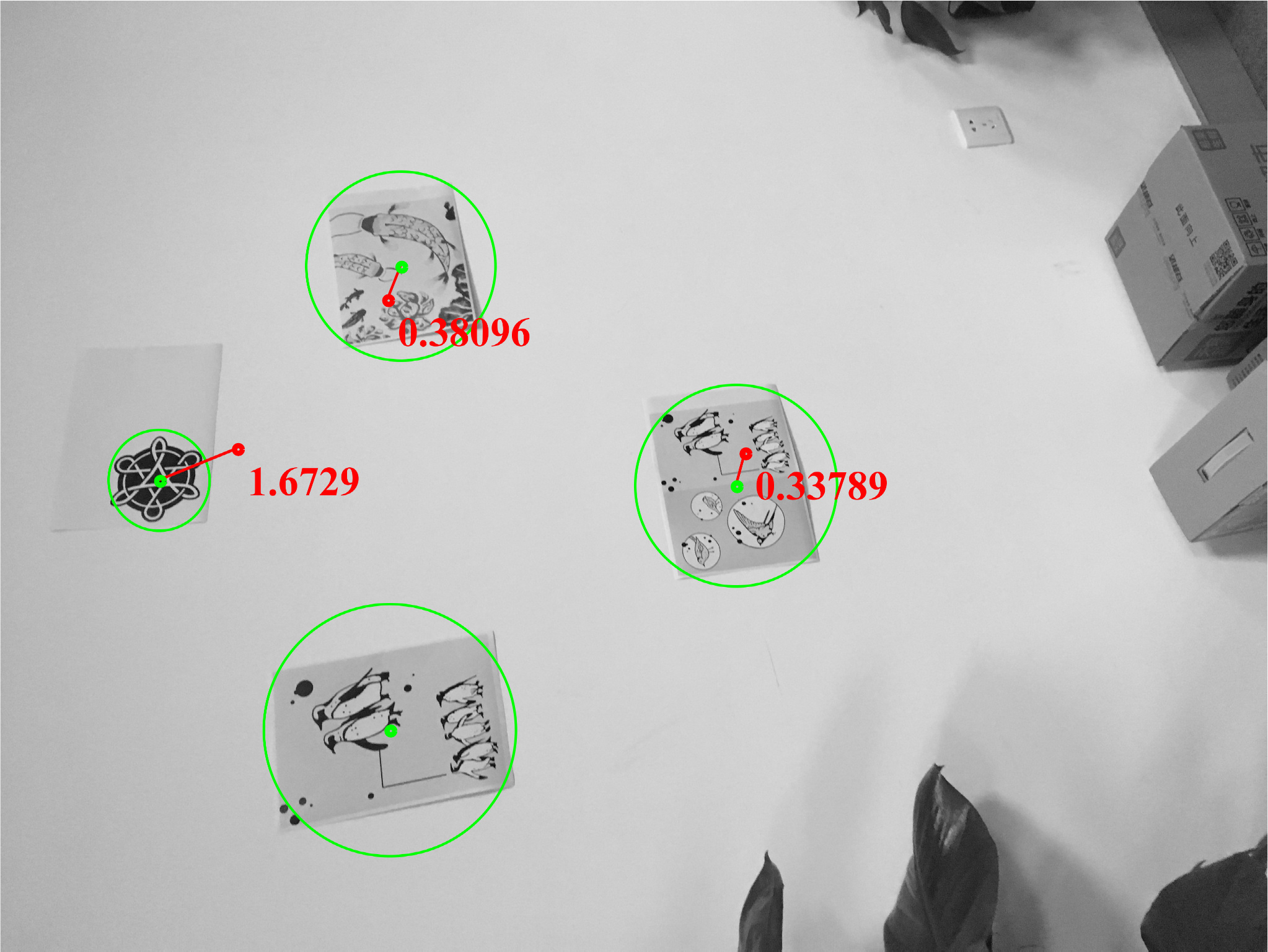}
    \end{subfigure} 
     \begin{subfigure}[t]{0.245\textwidth}
        \includegraphics[width=\textwidth]{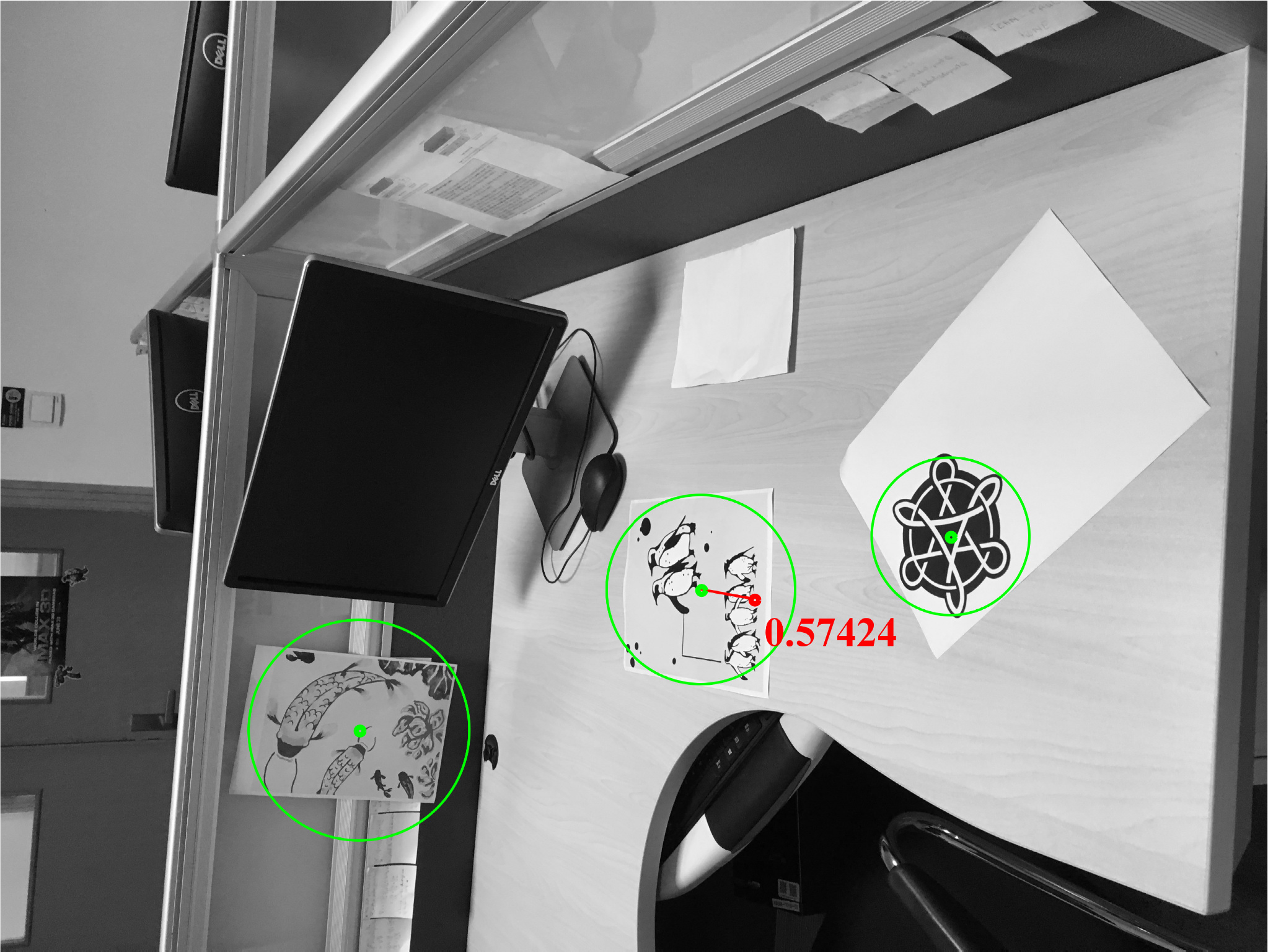}
    \end{subfigure} 
    \caption{Examples of successful Artcode localisation results.}
    \label{fig:lr-good}
\end{figure*}

\begin{figure*}
    \centering     
    \begin{subfigure}[t]{0.28\textwidth}
        \includegraphics[width=\textwidth]{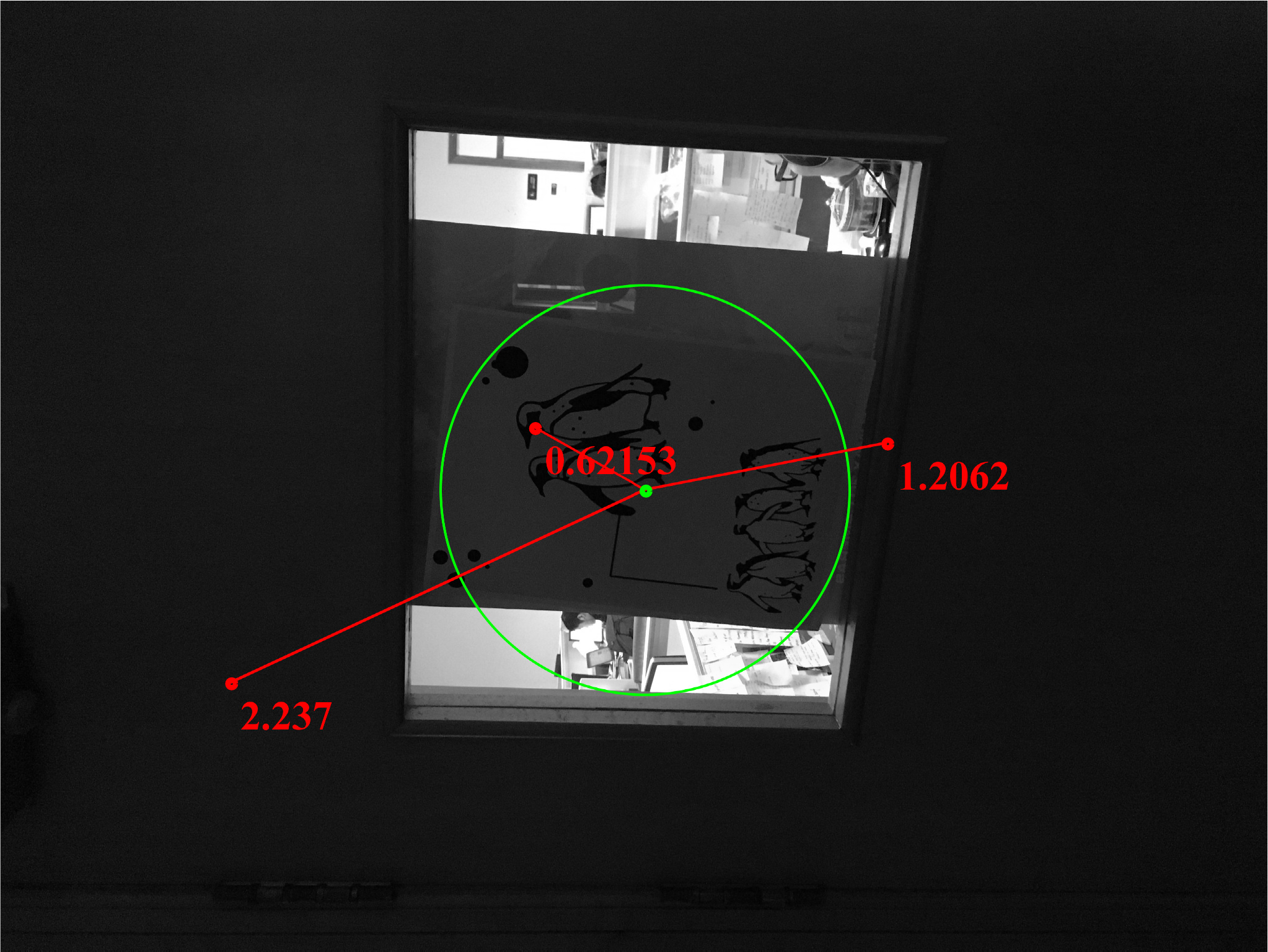}
        \caption{Dark lighting condition}
        \label{fig:lr-bad-a}
    \end{subfigure} 
     \begin{subfigure}[t]{0.28\textwidth}
        \includegraphics[width=\textwidth]{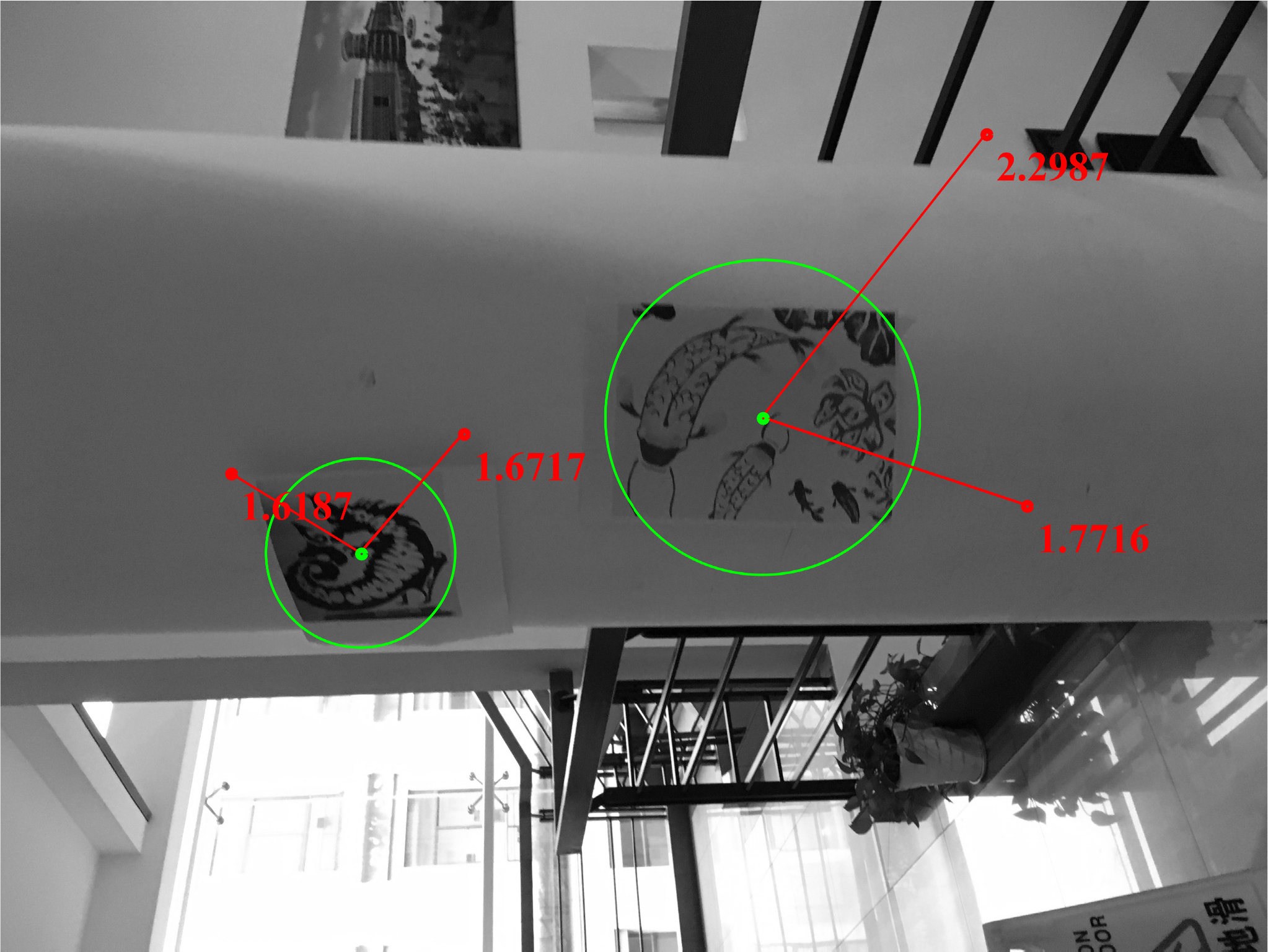}
        \caption{Shadows and sharp brightness contrast}
        \label{fig:lr-bad-b}
    \end{subfigure} 
    \begin{subfigure}[t]{0.21\textwidth}
        \includegraphics[width=\textwidth]{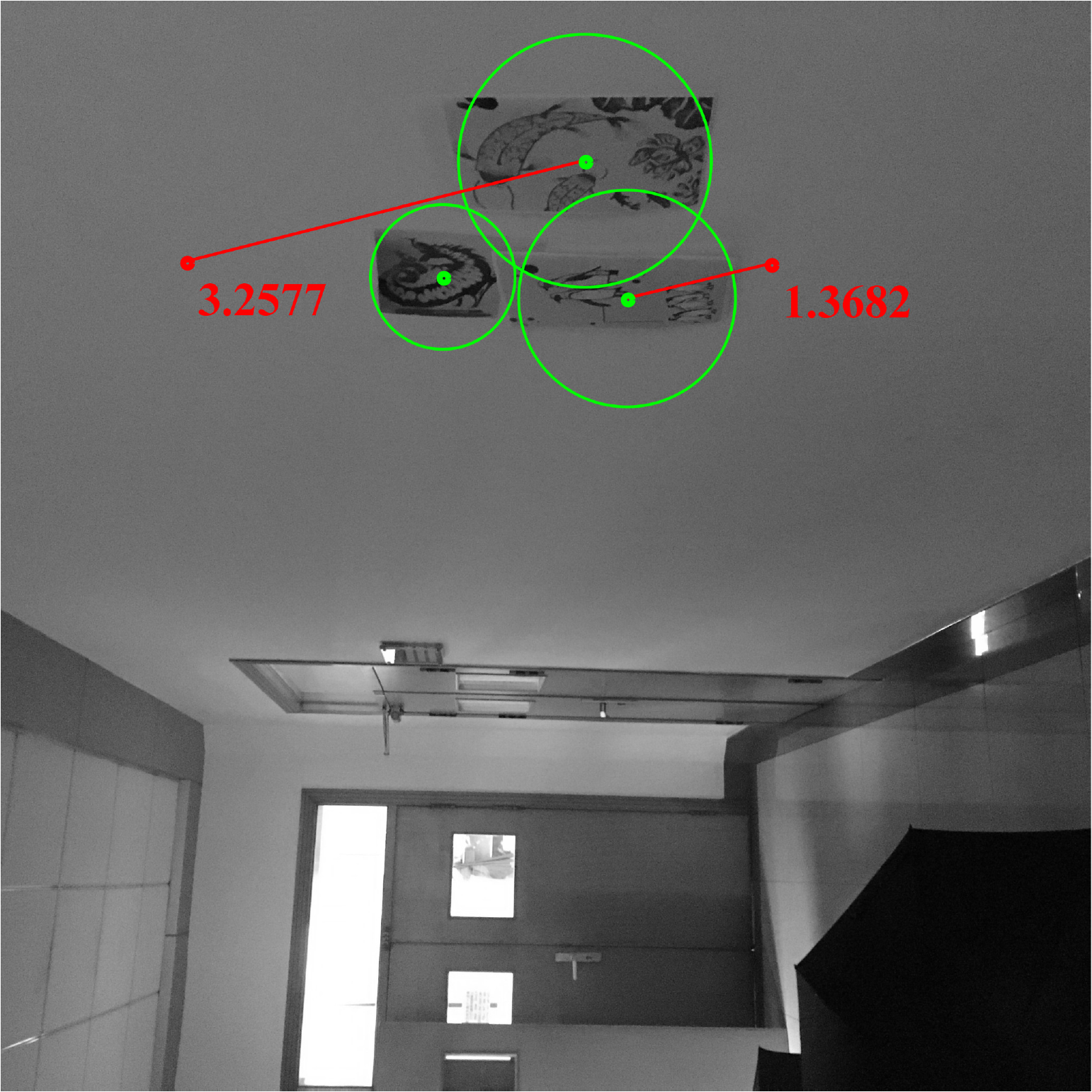}
        \caption{Extreme camera poses}
        \label{fig:lr-bad-c}
    \end{subfigure} 
    \begin{subfigure}[t]{0.21\textwidth}
        \includegraphics[width=\textwidth]{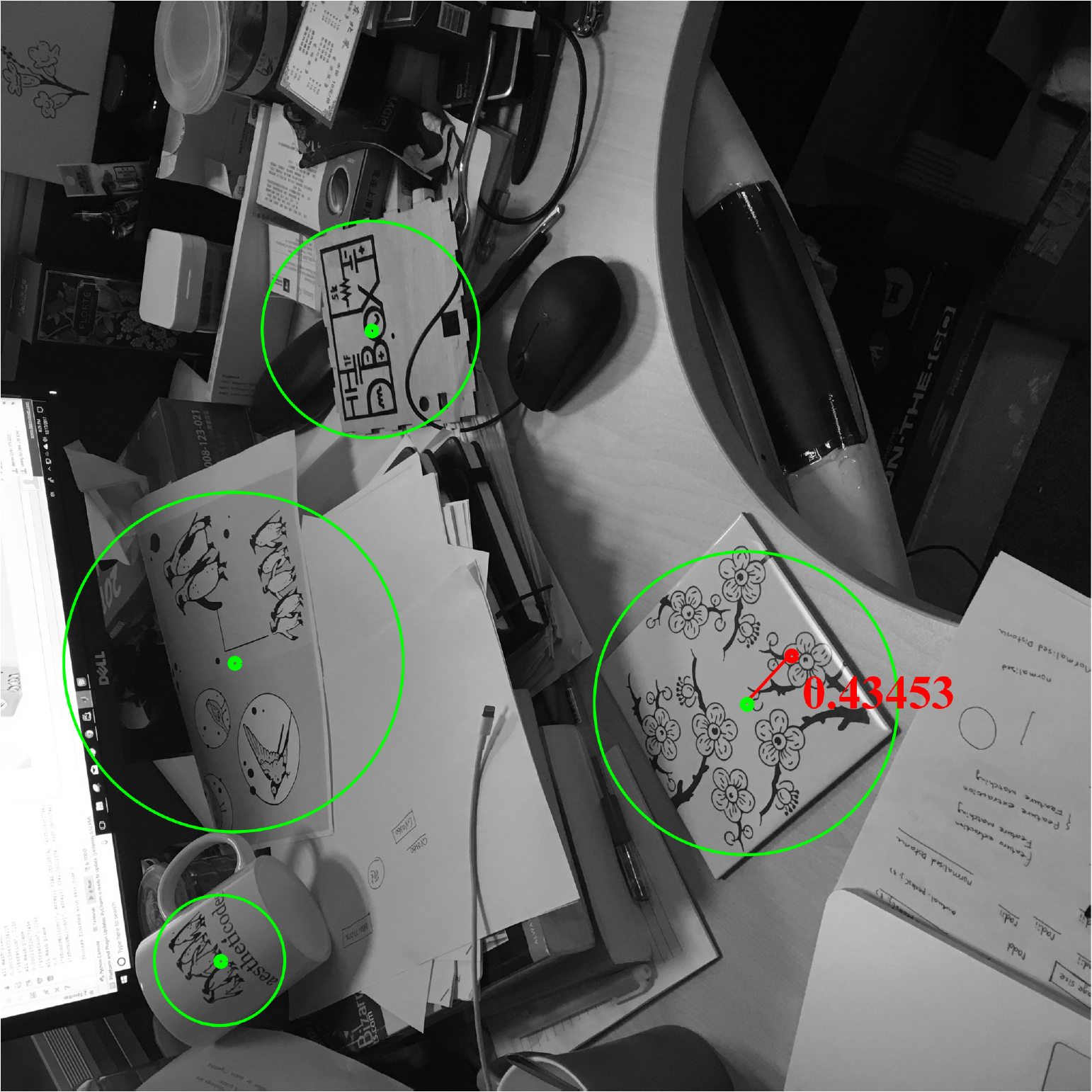}
        \caption{Cluttered background}
        \label{fig:lr-bad-d}
    \end{subfigure} 
    \caption{Examples of unsuccessful Artcode localisation results.}
    \label{fig:lr-bad}
\end{figure*}
While the proposed approach performs overall well on the given dataset across the four evaluation metrics, as evidenced in \autoref{subsec:results}, it has limitations when localising Artcodes in some challenging scenarios.
This section explores the localisation outcomes, examining potential factors that may influence these discrepancies.

We selected examples of both successful and unsuccessful localisations, which are visually represented in \autoref{fig:lr-good} and \autoref{fig:lr-bad}, respectively. 
These localisation results were generated by {\sc VisionGuide} using the following tuning parameters: $slidewin = 800 \times 800$, $step = 40$, $\sigma = 10$, $h = 10$. 
In both \autoref{fig:lr-good} and \autoref{fig:lr-bad}, the ground truth locations and sizes of the Artcodes are denoted by green points and circles, while the predicted locations are highlighted in red. 
The red lines connect the predicted and actual locations of an Artcode. 
Additionally, the red number adjacent to each line indicates the normalised distance between these locations. 
It is worth noting that an Artcode may have multiple predicted locations because of their freeform.

The localisation results are categorised into two groups: ``successful'' and ``unsuccessful,'' based prediction accuracy. 
In \autoref{fig:lr-good}, all Artcodes in the successful group were accurately localised, with predicted locations falling within the range of one normalised distance from the actual Artcode locations. 
Some predictions even fall within the range of normalised distance 0.5. 
In contrast, examples in \autoref{fig:lr-bad} exhibit a high number of false predictions that are far away from the actual locations (as seen in \autoref{fig:lr-bad-a} and \autoref{fig:lr-bad-b}), as well as undetected Artcodes (see \autoref{fig:lr-bad-c} and \autoref{fig:lr-bad-d}).

We identified three main factors affecting the localisation performance of {\sc VisionGuide}: lighting conditions, Artcode positions, and deformation. 
As illustrated in \autoref{fig:lr-good}, scenarios in this successful group have good lighting conditions, with well-lit and uniform illumination. 
Moreover, Artcodes were positioned on surfaces with good placements and minimal deformations.
Conversely, the Artcodes in \autoref{fig:lr-bad} were situated in environments with challenging conditions, including poor lighting conditions, out-of-plane deformation, extreme camera poses, and cluttered backgrounds. 
Thus, {\sc VisionGuide} performs significantly better in scenarios presented in \autoref{fig:lr-good} than in \autoref{fig:lr-bad}.
Although multiple factors affect the localisation simultaneously, lighting conditions emerge as one of the critical ones. 
Therefore, it is essential to consider lighting conditions carefully when designing Artcode-enabled access points for the metaverse in real-world applications.

\section{Conclusion and Future Work}
\label{sec:conclusion}
In this paper, we have addressed the problem of facilitating connections with virtual worlds in our everyday environments. 
Our focus has been on exploring the Artcode-enabled approach, which facilitates interactions with the metaverse through the interactive surface decorations of everyday objects \cite{meese2013codes, benford2017crafting}.
Specifically, we have presented a vision-based coarse-to-fine approach aimed at progressively detecting access points to the metaverse within our ambient surroundings. 
This two-stage approach allows for the creation of interaction affordances provided by freeform, unobtrusive, and scannable surface patterns such as Artcodes, an area that has not been extensively explored previously. 
Extensive experiments were conducted to evaluate the performance of this approach, which included the creation of a new dataset simulating the deployment of Artcodes in real-world environments.
Experimental results show that our approach achieves overall good performance across all four evaluation metrics, highlighting its effectiveness in localising Artcodes in various contexts. 
Additionally, we examined the limitations of our approach to provide insights for future research.

As an interdisciplinary topic, this work will be of interest to researchers and practitioners in the fields of Interaction Design, Mixed Reality, or Gaming who are looking for approaches to augment everyday public environments with virtual materials in an exploratory, gamified manner. 
Our future work will include further improving the performance of the proposed approach and conducting user studies to understand user experience when using this approach.

\section*{Acknowledgments}
\blackout{
The authors acknowledge the financial support from the Artificial Intelligence and Optimisation Research Group (AIOP), the Faculty of Science and Engineering (FoSE), the International Doctoral Innovation Centre, Ningbo Education Bureau, Ningbo Science and Technology Bureau, and the University of Nottingham.
}

\IEEEtriggeratref{18}
\bibliographystyle{ieeetr}
\bibliography{references}

\end{document}